%% file: main.tex
\documentclass[sigconf]{acmart}
\AtBeginDocument{%
  }

\copyrightyear{2026}
\acmYear{2026}
\setcopyright{cc}
\setcctype{by}
\acmConference[ICSE '26]{2026 IEEE/ACM 48th International Conference on Software Engineering}{April 12--18, 2026}{Rio de Janeiro, Brazil}
\acmBooktitle{2026 IEEE/ACM 48th International Conference on Software Engineering (ICSE '26), April 12--18, 2026, Rio de Janeiro, Brazil}
\acmPrice{}
\acmDOI{10.1145/3744916.3787837}
\acmISBN{979-8-4007-2025-3/2026/04}

\usepackage{alltt}
\usepackage{microtype}
\usepackage{graphicx}
\usepackage{subcaption}
\usepackage{booktabs} 
\usepackage{multirow}
\usepackage{soul}
\include{todonotes}
\usepackage{hyperref}
\usepackage{tikz}
\usepackage{enumitem}
\usepackage{xcolor}

\usepackage{natbib}
\usepackage{graphicx}
\usepackage{subcaption}

\usepackage{cuted}
\definecolor{Gray}{gray}{0.3}
\tikzstyle{mybox} = [draw=black, very thick, rectangle, rounded corners, inner ysep=5pt, inner xsep=5pt, fill=gray!20]

\newcommand*{\failtopass}[0]{\mbox{$F\!\!\to\!\!P$}\xspace}
\newcommand*{\oursystem}[0]{\mbox{e-Otter++}\xspace}

\newcommand{\findings}[2]{
    \smallskip
    \noindent
    \begin{tikzpicture}
        \node [mybox] (box){%
        \centering
        \begin{minipage}{.95\columnwidth}
        \fontsize{8.8}{10}\selectfont
        \textbf{Finding #1}. #2
        \end{minipage}
        };
    \end{tikzpicture}%
}

\usepackage{xspace}

\usepackage[capitalize,noabbrev]{cleveref}
\usepackage[ruled,vlined,linesnumbered]{algorithm2e}

\begin{document}

\title{Heterogeneous Prompting and Execution Feedback for SWE Issue Test Generation and Selection}


\author{Toufique Ahmed, Jatin Ganhotra, Avraham Shinnar, and Martin Hirzel}
\affiliation{%
  \institution{IBM Research}
  \city{Yorktown Heights}
  \state{New York}
  \country{USA}
  \postcode{10603}}
\email{tfahmed@ibm.com, {jatinganhotra, shinnar, hirzel}@us.ibm.com}

\begin{abstract}
\input{abstract}
\end{abstract}



\maketitle

\input{intro.tex}
\input{background.tex}
\input{method.tex}

\input{result.tex}

\input{discussion.tex}

\input{threats.tex}
\input{relatedwork.tex}
\vspace{-5pt}
\input{conclusion.tex}

\begin{acks}
We gratefully acknowledge Ibragim Badertdinov and Chunqiu Steven Xia for sharing SWE-rebench and Agentless patches.
\end{acks}

\balance
\bibliographystyle{ACM-Reference-Format}
\bibliography{bibfile}

\end{document}

%% file: abstract.tex
A software engineering issue (SWE issue) is easier to resolve when
accompanied by a reproduction test.
Unfortunately, most issues do not come with functioning reproduction
tests, so this paper explores how to generate them automatically.
One technique that helps with this task is inference scaling, but
traditional temperature-based scaling tends to lack diversity.
This paper introduces heterogeneous prompting to address this problem.
Another technique that helps with this task is execution feedback, but
this is hampered by the fact that the new code (after issue
resolution) to execute does not yet exist.
This paper introduces novel approaches to get around that problem.
We implemented our techniques in a new reproduction test
generator called \oursystem.
Experiments show that \oursystem represents a leap ahead in the
state-of-the-art for this problem, generating tests with an average
fail-to-pass rate of 63\% on the TDD-Bench Verified benchmark.

%% file: intro.tex
\section{Introduction}

Software engineering (SWE) \emph{issues} are descriptions of a
bug or missing feature.
SWE issues are usually informal and lack automated tests.
But such tests, when they exist, are invaluable for software
development.
Before issue resolution, tests can \emph{reproduce} the issue to confirm
whether it is real, make it more concrete, and help stakeholders agree
upon the desired behavior---following the highly regarded software
development practice of test-driven development~(TDD)~\cite{beck_2002}.
Once a code change~(a \emph{patch}) is proposed to resolve the issue,
running the same reproduction tests helps decide whether to
accept the patch.
Finally, after issue resolution, adding the tests to the
project's test suite helps prevent future regressions.
Reproduction tests for SWE issues should be \emph{fail-to-pass}~(\failtopass): they
should fail on the old (pre-patch) code to reproduce the issue and
pass on the new~(post-patch) code to accept the patch.
Unfortunately, generating \failtopass tests is challenging, because
issue descriptions are informal and because the new code patch on which the
tests should pass does not yet exist.
In addition to helping human software engineers, generating tests from
SWE issues can also assist automated SWE agents, as several SWE agents
also rely on reproduction tests~\cite{arora_et_al_2024,chen_et_al_2024,ehrlich_et_al_2025,jain_et_al_2025,li_et_al_2025,ruan_zhang_roychoudhury_2024,xia_et_al_2025,yang_et_al_2024,zhang_et_al_2024}.

There have been several recent attempts at using large language
models~(LLMs) to automatically generate \failtopass tests from SWE
issues, without access to the resolving code patches.
Libro uses an LLM to directly generate 50~tests from Java bug reports,
then performs rule-based postprocessing and ranking~\cite{kang_yoon_yoo_2023}.
Other attempts focus on Python, modifying a SWE agent designed to
automatically generate code patches to generate tests
instead~\cite{lin_et_al_2024,mundler_et_al_2024}.
Recent papers proposed novel LLM-based flows or agents custom-tailored
to the problem of generating tests from issues, including
AEGIS~\cite{wang_et_al_2024}, Otter++~\cite{ahmed_et_al_2025}, BRT
agent~\cite{cheng_et_al_2025}, AssertFlip~\cite{khatib2025assertflip}, and Issue2Test~\cite{nashid_et_al_2025}. 
Table~\ref{tbl:position} shows how \oursystem is procedurally distinct. 
Note that Otter, Issue2Test, and e-Otter do not use inference scaling or test selection.

There are two benchmarks for generating tests from issues:
SWT-bench~\cite{mundler_et_al_2024}~(derived from
SWE-bench~\cite{jimenez_et_al_2024}) and TDD-Bench
Verified~\cite{ahmed_et_al_2024-tdd}~(derived from SWE-bench
Verified~\cite{chowdhury_et_al_2024}).
SWT-bench Verified, also derived from SWE-bench Verified, is very
similar to TDD-Bench Verified.
SWT-bench has a leaderboard~(\url{https://swtbench.com}), topped
(as of when we submitted this paper) by
the unpublished Amazon Q Developer Agent with an \failtopass rate of
37.7\% on SWT-bench Lite and 49.0\% on SWT-bench Verified.
However, as there is no paper about Amazon Q Developer Agent, its
inner workings are not publicly known.
The top system on SWT-bench Verified with a published paper is
Otter++~\cite{ahmed_et_al_2025} with an \failtopass rate of~37.0\%, and
it also has the exact same \failtopass rate of 37.0\% on TDD-Bench
Verified.
Given how similar the two verified benchmarks are, this paper
evaluates on TDD-Bench Verified only.
This paper establishes a new state of the art (SOTA) for
SWT-bench Lite with 52.5\% \failtopass rate and for
TDD-Bench Verified with 63.0\% \failtopass rate.
It does so with a new system, \oursystem. The ``e-'' refers to
execution feedback: \oursystem outperforms prior work by leveraging
execution feedback more effectively.
The other core innovation in \oursystem is heterogeneous prompting.
Additionally, we apply our approach
to the contamination-free benchmark SWE-rebench~\cite{badertdinov2025swe} to show the generalizability of our innovations,
as SWE-bench is suspected to be exposed to recent frontier models during training~\cite{liang2025swe}.

The first place where \oursystem uses execution feedback is after
generating a candidate test.
The test should fail on the old code, which is easy to
accomplish, but it should fail for the right reason, which is
challenging.
Unlike prior work, \oursystem uses execution feedback to retrieve
related code and augments the repair prompt with that.
The second place where \oursystem uses execution feedback is for test
selection.
After generating and repairing multiple tests, it automatically picks
a single final test.
The chosen test should pass on the new code, but since the resolving
new code is not available, there is no execution feedback to tell
whether that is the case.
Unlike prior work, \oursystem generates candidate code patches as a
surrogate for the hidden ground-truth new code.
Despite generated code patches being of course imperfect, our
experiments show that tests passing on them helps predict
passing on the ground-truth code patch.
However, even the best test selector is only as good as the pool of
generated candidate tests it picks from.
Ideally, this test pool should be diverse, and most prior work turns
up the LLM decoding temperatures to increase candidate diversity.
Unlike prior work, \oursystem instead gets diversity from
\emph{heterogeneous prompting}. It uses two kinds of
systematic prompt variations: issue description \emph{morphs}, which
use an LLM to rewrite the issue text in various ways before test
generation, and prompt context \emph{masks}, which elide various
subsets of features from the test-generation prompt.
Our experiments show that morphs and masks greatly increase the
likelihood of at least one candidate test being \failtopass.

This paper makes the following novel contributions:
\begin{itemize}
  \item Generating reproduction tests from issues and increasing their
    correctness via execution-augmented test repair.
  \item Increasing reproduction test candidate diversity via issue
    description morphs and test-generation context masks.
  \item Selecting a single final reproduction test via execution
    feedback from imperfect code patches.
\end{itemize}

Taken together, these novel contributions enable \oursystem to
substantially advance the SOTA for generating reproduction tests from
issues.
Ultimately, we hope our work helps cement the practice of using tests
both to drive issue resolution and for long-term
continuous quality assurance after.

\begin{table}[t]
\centering
\caption{Component-wise comparison with prior works}
\resizebox{\columnwidth}{!}{%
\renewcommand{\arraystretch}{1.2}
\begin{tabular}{l|ccc|ccc|ccc}
\toprule  
\multicolumn{1}{c|}{\multirow{2}{*}{System}} & \multicolumn{3}{c|}{Test Generation}                                                                                                 & \multicolumn{3}{c|}{Inference Scaling}                                         & \multicolumn{3}{c}{Test Selection}                                                                                                                  \\ \cline{2-10}
\multicolumn{1}{c|}{}                        & \begin{tabular}[c]{@{}c@{}}Execution \\ Feedback\end{tabular} & \begin{tabular}[c]{@{}c@{}}Context \\ Augment\end{tabular} & Critic & \begin{tabular}[c]{@{}c@{}}Restart with \\ Temp 0\end{tabular} & Mask & Morph & \begin{tabular}[c]{@{}c@{}}Execution \\ on $c_\textrm{old}$\end{tabular} & \begin{tabular}[c]{@{}c@{}}Execution on \\ Surrogate Patch\end{tabular} & Coverage \\ \midrule
Otter                                       & No                                                            & No                                                         & No     & \multicolumn{3}{c|}{Not Applicable}                                            & \multicolumn{3}{c}{Not Applicable}                                                                                                                  \\
Otter++                                     & No                                                            & No                                                         & No     & No                                                             & Yes  & No    & Yes                                                            & No                                                                      & No       \\
Isue2Test                                   & Yes                                                           & No                                                         & Yes    & \multicolumn{3}{c|}{Not Applicable}                                            & \multicolumn{3}{c}{Not Applicable}                                                                                                                  \\
AEGIS                                       & Yes                                                           & No                                                         & Yes    & Yes                                                            & No   & No    & Yes                                                            & No                                                                      & No       \\
e-Otter                                     & Yes                                                           & Yes                                                        & Yes    & \multicolumn{3}{c|}{Not Applicable}                                            & \multicolumn{3}{c}{Not Applicable}                                                                                                                  \\
e-Otter++                                   & Yes                                                           & Yes                                                        & Yes    & No                                                             & Yes  & Yes   & Yes                                                            & Yes                                                                     & Yes     \\ \bottomrule
\end{tabular}
}
\label{tbl:position}
\vspace{-15pt}
\end{table}

%% file: background.tex
\section{Problem Statement and Background}
\label{background}

This section discusses the problem statement and briefly introduces prior
work we build upon: Otter~\cite{ahmed_et_al_2025},
Agentless~\cite{xia_et_al_2025}, SWT-bench~\cite{mundler_et_al_2024},
and TDD-Bench Verified~\cite{ahmed_et_al_2024-tdd}.

\begin{figure}[t]
  \centerline{\includegraphics[width=.8\columnwidth]{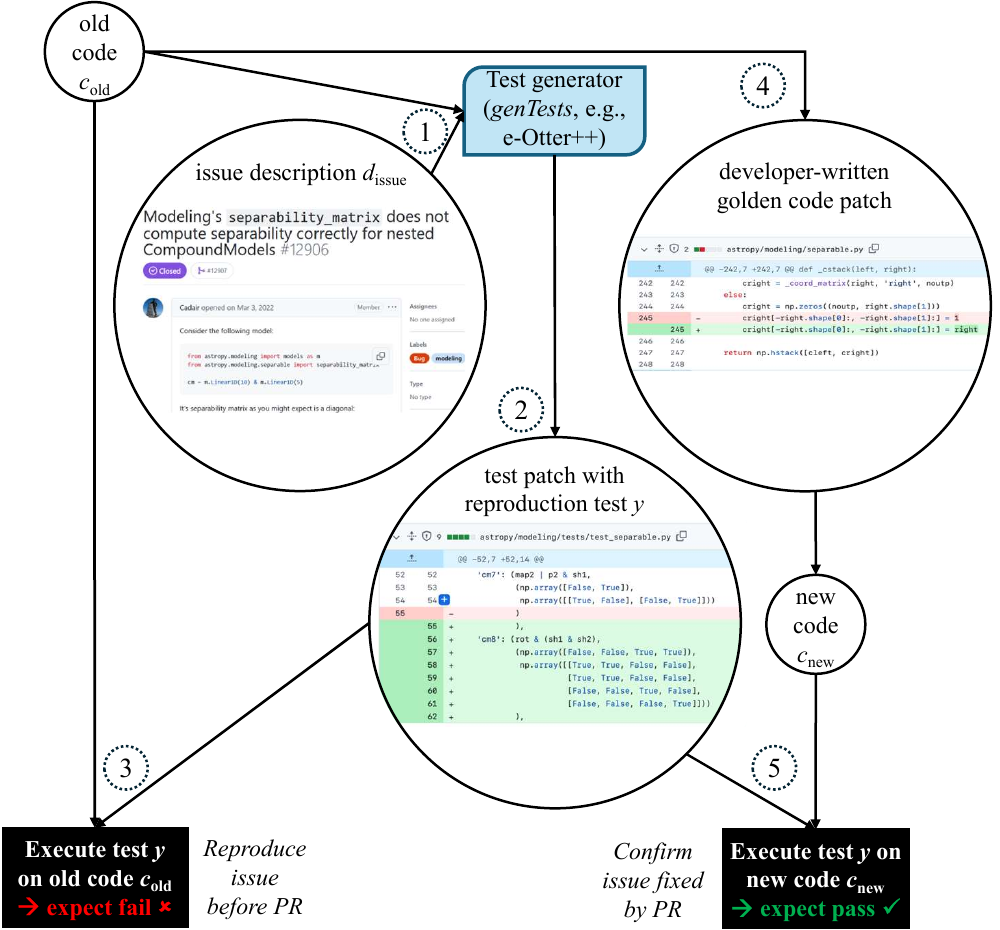}}
  \caption{\label{fig:harness} Evaluation harness for bug reproduction test. 
  (1)~The issue description and $c_\textrm{old}$ go into \oursystem as input, 
  (2)~\oursystem generates a test patch with a reproduction test~$y$,
  (3)~executing the test $y$ on $c_\textrm{old}$ should fail in order to reproduce the issue, 
  (4)~a developer-written golden code patch is applied to $c_\textrm{old}$ as a pull request resulting in $c_\textrm{new}$, 
  (6)~executing the test $y$ on $c_\textrm{new}$ should pass now in order to confirm that the issue has been addressed by the pull request.}
\vspace{-.1cm}  
\end{figure}

\subsection{Problem Statement}
\label{ps}
This paper addresses the challenge of generating reproduction tests
directly from issue descriptions.
\cref{fig:harness} presents an evaluation harness for reproduction tests.
The input $x$ consists of two parts: an issue description~$d_\textrm{issue}$
(usually written in natural language and possibly including code
snippets and stack traces) and the existing version of the
codebase~$c_\textrm{old}$ (before any fix is applied).
Crucially, a solution for this problem statement does not have access
to the updated codebase~$c_\textrm{new}$, reflecting real-world conditions.
The goal is to produce a set of tests $y$ that will fail on $c_\textrm{old}$, 
thereby confirming the presence of the issue, and pass on $c_\textrm{new}$, 
confirming that the issue has been resolved
(in other words, $y$~should be~\failtopass).
These tests should aim to cover the relevant changes made in the code.
The problem is to develop a function $\textit{genTests}$ that takes as
input only \mbox{$x=\langle d_\textrm{issue},c_\textrm{old}\rangle$} and
returns the generated tests $y$ as output.

\subsection{Reproduction Test Generation: Otter}
\label{otter}
This paper introduces \oursystem, which extends an earlier system
called Otter++~\cite{ahmed_et_al_2025}.
Both \oursystem and Otter++ generate tests directly from issue
descriptions and $c_\textrm{old}$.
In other words, both are functions $\textit{genTests}$ that address
the problem stated in~\cref{ps}.

At the heart of Otter++ is Otter, which is a pipeline of LLM-based and
rule-based steps.
Otter comprises three key components: a localizer, a self-reflective
action planner, and a test generator.
The localizer identifies relevant files, test functions, and focal
functions by querying the LLM with context-aware prompts.
The action planner then iteratively builds a plan of actions—categorized as read, write, and modify—guided by a self-reflective loop and validation checks. 
Finally, the test generator produces complete test functions using localized code and structural cues like test signatures and imports; it also includes mechanisms for repairing hallucinated or incomplete imports and integrates static analysis tools (e.g., the Flake8 Python linter) to ensure syntactic and semantic correctness.

Otter++ improves Otter by generating multiple test candidates using heterogeneous prompts, each varying in the inclusion of localized context. 
It selects the best candidate by analyzing execution outcomes (e.g., assertion failures vs.\ syntax errors) on the original code~$c_\textrm{old}$.
Empirical results using two models (GPT-4o and Mistral-large) show that Otter++ achieves higher \failtopass rates than prior systems, while remaining cost-efficient and generalizable.

Our new system \oursystem uses Otter as a base component and adds
several new techniques, discussed in detail in \cref{sec:method}.

\subsection{Code Patch Generation: Agentless}

One of the new techniques in \oursystem is how it uses execution feedback
from automatically generated candidate code patches.
Rather than implementing a new code patch generator, it reuses an
existing one, Agentless~\cite{xia_et_al_2025}, as a component.

Agentless implements a hierarchical bug localization process. 
First, it identifies suspicious files using both LLM prompting and embedding-based retrieval (filtered to avoid irrelevant files). 
It then narrows down to relevant classes/functions using a concise skeleton representation of files.
Finally, it pinpoints exact edit locations with line-level granularity. 
For repair, it generates multiple patches in a lightweight Search/Replace diff format rather than rewriting entire code blocks. 
This minimizes hallucinations and boosts cost-efficiency. 
During patch validation, Agentless automatically generates reproduction tests (when none are provided) and combines them with regression tests to select the best patch via majority voting. 
This modular structure allows Agentless to stay interpretable, easily debuggable, and effective without relying on complex environment interaction or planning mechanisms typical of agents.

Our new system \oursystem uses only the first part of Agentless,
\emph{without} the test generation or patch selection of Agentless,
as a component~(see \cref{sec:patchgen}).
While we could have adopted any code patch generator, we chose
Agentless because it is open-source and candidate patches are publicly
available~\cite{agentless_artifact}.

\subsection{Test Evaluation Benchmarks}
There are two benchmarks for evaluating bug reproduction tests.
SWT-bench Lite~\cite{mundler_et_al_2024} was derived primarily from the SWE-bench Lite dataset~\cite{jimenez_et_al_2024} and contains 276 samples.
TDD-Bench Verified~\cite{ahmed_et_al_2024-tdd} was derived from SWE-bench Verified~\cite{chowdhury_et_al_2024}.
In TDD-Bench Verified, the authors ran 500 samples from SWE-bench Verified, filtered out samples with zero test coverage or those violating the \failtopass property of 
the tests, and ended up with 449 samples. 
Both benchmarks judge whether a generated test reproduces the described issue by checking whether the test fails before applying the golden patch and passes afterward.
In other words, they both implement an evaluation harness consistent with
the diagram in \cref{fig:harness}.
One difference between their implementations is that TDD-Bench only runs contributing tests (tests that were added or updated), 
whereas SWT-bench runs the whole test files that contain the contributing tests.
Later, M{\"u}ndler et al.\ introduced SWT-bench Verified, which also
derives from SWE-bench Verified like TDD-Bench, into their
leaderboard~(\url{https://swtbench.com}).
\cref{sec:evaluation} reports results on both SWT-bench Lite and
TDD-Bench Verified.

%% file: method.tex
\section{Methodology}
\label{sec:method}

\begin{figure*}[t]
  \centerline{\includegraphics[width=\textwidth]{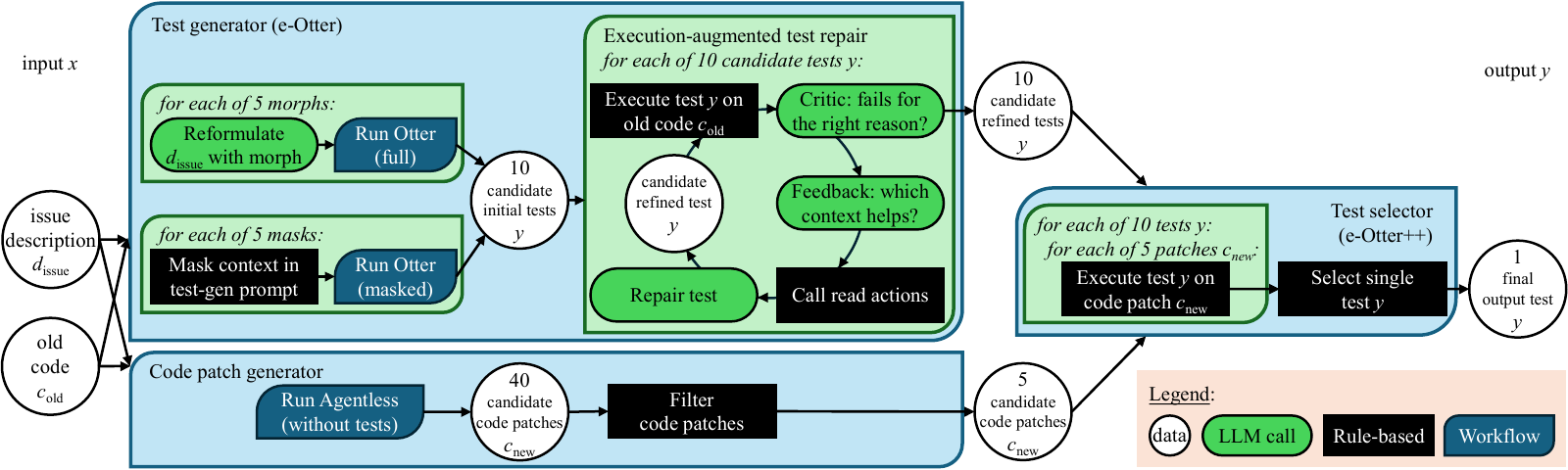}}
  \caption{\label{fig:overview} Overview of our approach.}
\end{figure*}

\cref{fig:overview} presents the overview of our approach.
It has three major components: test generator, code patch generator,
and test selector.
The following three subsections describe each component.

\subsection{Test Generator}\label{sec:testgen}
Our test generator, e-Otter, starts with tests generated by Otter~(see
\cref{otter}), then increases their correctness via execution-augmen\-ted
test repair.
It also increases reproduction test candidate diversity via issue
description morphs and test-generation context masks.

\begin{algorithm}[t]
\caption{\label{alg:repair}Execution-augmented test repair}
\small
\SetAlgoLined
\SetKwInOut{Input}{Input}
\SetKwInOut{Output}{Output}
\Input{issue description $d_\textrm{issue}$; old code $c_\textrm{old}$; candidate test $y$}
\Output{refined candidate test}
\Repeat{$\textit{failsForRightReason}(\textit{critique})$ \textbf{or} too many attempts}{
  $\textit{executionFeedback} \leftarrow \textit{executeTestOnCode}(y, c_\textrm{old})$\;
  $\textit{critique} \leftarrow \textit{LLM}(\textit{criticPrompt}(d_\textrm{issue}, \textit{executionFeedback}))$\;
  \If{\textbf{not} $\textit{failsForRightReason}(\textit{critique})$}{
    $\textit{reads} \leftarrow \textit{getActions}(\textit{critique})$\;
    $\textit{extraContext} \leftarrow [\textit{executeAction}(a,c_\textrm{old})\textbf{ for }a\in\textit{reads}]$\;
    $y \leftarrow \textit{LLM}(\textit{repairPrompt}(d_\textrm{issue},y,\textit{extraContext})$
  }
}
\Return $y$
\end{algorithm}

\paragraph{Execution-augmented Test Repair.}
As shown in Algorithm~\ref{alg:repair},
the test repair component starts by running an Otter-generated
candidate test~$y$ on~$c_{old}$.
It is expected that the test should fail in this phase.
However, a test may not fail for the reason mentioned in the issue description. Chen et al.~\cite{chen_ma_jiang_2025} reported that Python execution errors during the issue resolution phase correlate with lower resolution rates. 
They also identified the most prevalent errors—such as ModuleNotFoundError and TypeError—in the generated code, which tend to differ greatly from the reason mentioned in the issue description. 
To address this, we propose an LLM-based critic that will take the issue description and test log as input and decide whether the test is failing for the right reason. If the critic believes the test is failing for 
the reason mentioned in the issue description, the repair loop immediately stops.
Otherwise, it tries to repair the error with additional information
identified by the critic, which includes the possible buggy line, 
the most relevant text/code snippet from the issue description, and any function that the model wants to see before attempting to write the test again.
The critic consists of a single LLM call with temperature 0, where the LLM output contains both the decision and additional information. 
We explicitly ask for the buggy line because these Python execution errors (e.g., ModuleNotFoundError, TypeError) are localized to one particular line. It is helpful for the model to know which part of the program needs more attention.
We have also seen some cases where the issue description has enough hints to avoid execution errors. For example, in ``astropy\_\_astropy-14182'', the GPT-4o model generates ``ascii.write()'', which is wrong. 
The correct version is ``tbl.write()'', which is not as highly used as ``ascii.write()'' but is present in the issue description. The critic also proposes function names to read to augment the context (e.g., function arguments, return type), and that actually helps the model in repairing the tests.

After obtaining the additional information from the critic, the repair
loop collects the functions from the code that the LLM wants to see.
Next, it calls the LLM again to repair the test, feeding the issue
description, the current version of the test, the test log, and the
additional information collected based on the critic output into the
test-repair prompt.
Then, the repair loop starts over from this new test as~$y$,
until the critic is satisfied or for a maximum of 10~iterations.
Following prior works~\cite{xia_et_al_2025, ehrlich_et_al_2025,
  chen2021evaluating}, the test repair LLM call uses a higher
temperature~(0.8) to encourage the model to try diverse solutions.
\emph{We refer to the test generated at this phase as e-Otter test}.

\paragraph{Issue Description Morphs and Test-Generation Context Masks}
Recent works both on generating
SWE-patches~\cite{ehrlich_et_al_2025, xia_et_al_2025, jain_et_al_2025}
and reproduction tests~\cite{ahmed_et_al_2025, wang_et_al_2024}
have adopted inference scaling.
Most works rely on higher temperatures to increase the diversity of solutions. 
One of the key limitations of such approaches is using the same context for all solutions, which may contain misleading information~(e.g., incorrect localization).
Also, in a chain of multiple LLM calls, using a higher temperature in
each call exponentially increases the number of samples.

Instead of higher temperature, Otter++~\cite{ahmed_et_al_2025} attempted
heterogeneous prompting via five \emph{masks} hiding different parts of
contexts and achieved a higher fail-to-pass rate~@5.
We refer to the five tests from Otter++ masks as planner, full,
testLoc, patchLoc, and none.
Planner represents the default Otter-generated tests where self-reflective action planning was applied. 
The other masks partially or completely expose localization information to the model to generate diverse solutions.
For example, in `testLoc', the test generation prompt contains only
test localization outputs, but no patch localization or planner outputs.

Going beyond masks, this paper introduces another approach for
heterogeneous prompting, adding five issue description \emph{morphs}.
Each morph rewrites the issue with an LLM, then runs Otter on the
rewritten issues to generate diverse solutions.
The goal is to increase the fail-to-pass rate~@10.
The five morphs are:

\vspace{.2em}

\noindent i) \underline{standard:} This morph asks the LLM to standardize the issue description. 
A well-formatted issue description has several components, including a title, description of the problem, steps to reproduce, expected behavior, and actual behavior. 
Having these components may help the model to reason better or perform better in producing the bug reproduction test. 

\vspace{.2em}

\noindent ii) \underline{simple:} Some issue descriptions come with code artifacts and jargon that are very project-specific, making them hard to understand for a developer who does not work on the project regularly. 
We ask the LLM to simplify the issue so that it becomes easier to understand for developers.

\begin{figure}[t!]
    \centering
    \begin{subfigure}{.49\columnwidth}
        \centering
        \includegraphics[width=.95\columnwidth]{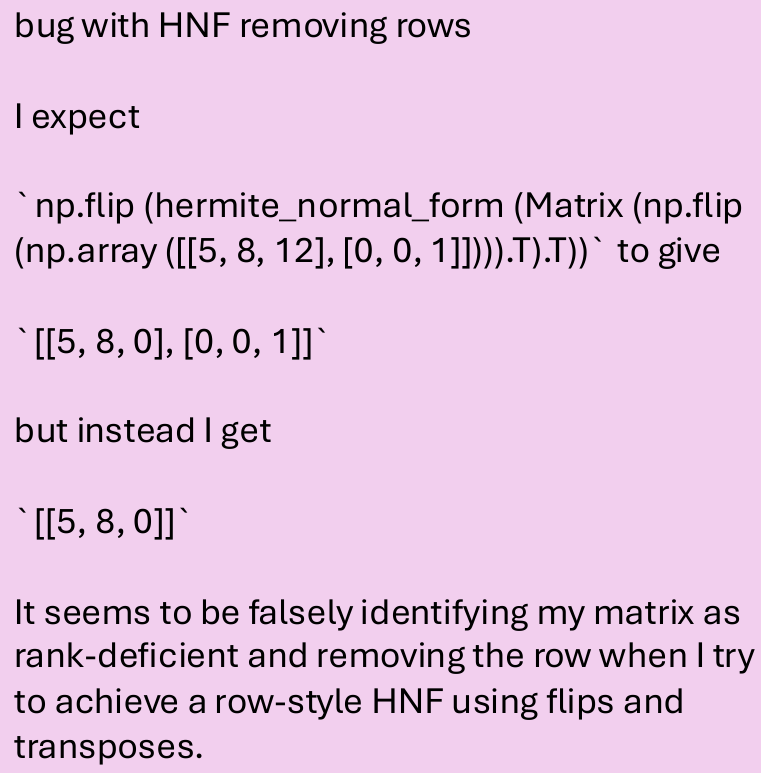}
        \caption{Issue description}
        \label{fig:sub1}
    \end{subfigure}
    \begin{subfigure}{.49\columnwidth}
        \centering
        \includegraphics[width=.95\columnwidth]{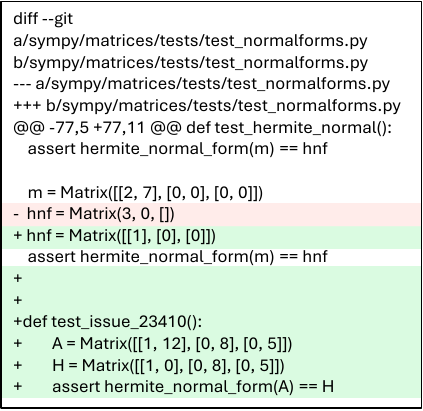}
        \caption{Test patch}
        \label{fig:sub2}
    \end{subfigure}
    \caption{Issue description and test patch for sympy\_\_sympy-23413. Some lines are omitted from the patch due to space constraints.}
    \label{fig:coverage}
\vspace{-10pt}    
\end{figure}

\vspace{.2em}

\noindent iii) \underline{dropCode:} We have seen some issue descriptions that come with misleading code snippets. 
Developers sometimes casually write snippets that may be useful for them, but are confusing to the model. 
For example, in~\cref{fig:sub1}, the issue description uses Numpy~(np) to manipulate a matrix. 
However, the Sympy project does not use or even install Numpy as a dependency for that particular issue. 
Seeing the original unmorphed issue, the model generates a test using np for matrix manipulation, which is not applicable in this setup.
Dropping this code snippet helps the model avoid this mistake.
The developer-written test in \cref{fig:sub2} does not use Numpy.

\vspace{.2em}

\noindent iv) \underline{initTest:} This morph asks the LLM to produce an initial fail-to-pass test and incorporate it into the issue description without giving any additional context. 
The rest of the pipeline may get some initial idea from the test and modify it so that it goes from fail to pass. Note that in this augmentation process, the model may contaminate the issue from a memorized test. 
We manually validate the possibility of contamination by the LLM and found that in 2\% of cases, we observed some possible contamination where the model 
might have used memorized data to resolve the problem (see details in~\cref{sec:discusion}). Also, we are not performing anything better than zero-shot prompting here, 
and we observed that zero-shot prompting is not as effective as other approaches. 

\vspace{.2em}

\noindent v) \underline{initPatch:} Like initTest, this morph asks the model to propose an initial solution to resolve the issue, which may guide the rest of the pipeline to perform well. 
We found that in practice, for initTest and initPatch, the models sometimes do not write an initial test or patch due to a lack of confidence. The model may simply rewrite the issue with minor modifications.

The motivation behind morphing is to change the issue such as to give the model a different perspective. 
Of course, this only happens if the human-written issue does not
already have the characteristics that the morph would add.
This raises the question how often that is the case; to answer it,
we manually labeled human-written issues with morph categories.
Upon reviewing the entire dataset (449 issues), only 28\% are written in a standard format, 
which means our standardization morphing makes substantial changes to the remaining 72\%. 
For other morphing techniques, substantial changes happen to 28\%~(simple), 
71\%~(dropCode), 47\%~(initTest), and 90\%~(initCode) issues.  
There is certainly an opportunity to add even more diversity, but~\cref{temperature} shows diminishing returns.

We rerun the Otter pipeline using these new morphed issue descriptions and generate five new sets of tests. 
Note that issue rewriting may hurt performance since it may remove necessary information and add irrelevant context. 
However, our goal is to increase the \failtopass rate @ N, which could be achieved by such rewriting.
Prompts and examples for execution-augmented test repair and issue description morphs are in our repository~\cite{eotter_artifact}.

\subsection{Code Patch Generator}\label{sec:patchgen}

Our approach uses generated code patches to help with test selection.
Prior work does something similar but in the other direction,
using generated tests to help with code selection.
CodeT~\cite{chen_et_al_2023} generates both code samples and
corresponding test cases, then executes the generated code samples
using the generated test cases and performs a dual execution agreement
to choose the best code sample.
Agentless~\cite{xia_et_al_2025} and later CodeMonkeys~\cite{ehrlich_et_al_2025}
also use reproduction tests to select patches from multiple candidates
but do not evaluate the quality of the reproduction test.
Our \oursystem uses code patches to evaluate the quality of generated
reproduction test candidates and attempt to choose the best one.
We use Agentless patches, as opposed to patches from other systems,
because their artifacts are open-source and their
generated patches are available.
Agentless generates a total of 40 code patches (1~greedy at
temperature~0 and 39 non-greedy at \mbox{temperature 0.8})
prior to patch selection.
We take the normalized version of these 40 patches and select the
5~most frequent ones.
We chose 5 considering the number of existing normalized unique
patches and our resource constraints.

\subsection{Test Selection}\label{sec:selection}

The \oursystem test selector starts from 10~candidate refined tests
and 5~candidate code patches, and its job is to pick 1~single final
output test.
The test selector is entirely rule-based and does not call any LLM.
It starts by running each of the candidate tests both on
$c_\textrm{old}$ and on each of the code patches~$c_\textrm{new}$.
If none of the tests are fail-to-pass on any code patch, then the test
selector keeps all 10~candidate tests until the tie-breaking step,
otherwise it:
\begin{itemize}[topsep=0pt]
  \item Keeps only tests that are fail-to-pass.
  \item Classifies those fail-to-pass tests into three groups based on how
    they failed on $c_\textrm{old}$:
    \begin{itemize}
      \item assertion failure (these are most likely to have failed
        for the right reason),
      \item other failure (the test ran but produced the wrong output), or
      \item error (the test did not run properly, e.g., because of
        wrong syntax or an exception in a fixture; least likely to
        have failed for the right reason).
    \end{itemize}
  \item Keeps only the tests in the first non-empty of those three groups.
\end{itemize}
At this point, the test selector uses coverage as a tie-breaker for
the remaining tests, by picking one test with the highest coverage.
The coverage of a test~$y$ is the average of its coverage on the
5~code patches.
The coverage of a test~$y$ on a code patch~$c_\textrm{new}$ is
the sum of the number of covered lines deleted from~$c_\textrm{old}$ plus
the number of covered lines added by~$c_\textrm{new}$,
divided by the total added+deleted lines.
\emph{We refer to the test selected at this phase as \oursystem test}.

%% file: result.tex
\section{Evaluation}\label{sec:evaluation}

This section describes experiments to answer the following research
questions:

\begin{itemize}[leftmargin=2em]

\item \textbf{RQ1:} How does our approach perform at generating \failtopass tests?
\item \textbf{RQ2:} How effective is execution-augmented test repair?
\item \textbf{RQ3:} How do issue description morphs and test-generation context masks increase the \failtopass @ N?
\item \textbf{RQ4:} How do generated code patches affect test selection?
\item \textbf{RQ5:} How does our approach help in validating code patches?
\item \textbf{RQ6:} How effective is \oursystem with a contamination-free benchmark?

\end{itemize}  

We primarily evaluate our approach on two benchmarks: TDD-Bench Verified\footnote{\url{https://github.com/IBM/TDD-Bench-Verified}} and SWT-bench Lite\footnote{\url{https://github.com/logic-star-ai/swt-bench}}. 
Both benchmarks are derived from the popular issue resolution benchmark, SWE-bench~\cite{jimenez_et_al_2024}.
We report the fail-to-pass (\failtopass) rate to show the effectiveness of the approach following prior works~\cite{mundler_et_al_2024,nashid_et_al_2025, ahmed_et_al_2025}. 
We have primarily used two models, GPT-4o and Claude-3.7-Sonnet.
To have a fair comparison with Otter and Otter++, we used the same version of the GPT-4o model (gpt-4o-2024-08-06) as in that paper~\cite{ahmed_et_al_2025}.
We also report results on SWE-rebench~\cite{badertdinov2025swe} with Claude-3.7-Sonnet to demonstrate its effectiveness on a contamination-free benchmark.

\subsection{Performance of e-Otter and \oursystem (RQ1)}

\begin{table}[t]
\centering
\caption{Performance of e-Otter, \oursystem, and baselines}
\resizebox{.85\columnwidth}{!}{%
\renewcommand{\arraystretch}{1.2}
\begin{tabular}{llrr}
\toprule    
\multicolumn{1}{c}{Benchmark}                                                                              & \multicolumn{1}{c}{Approach}                      & \multicolumn{1}{c}{\failtopass}               & \multicolumn{1}{c}{In (\%)}              \\ \midrule
\multirow{9}{*}{\begin{tabular}[c]{@{}l@{}}TDD-Bench \\Verified (449 \\samples)\end{tabular}}                          & Zero-shot (GPT-4o)                                           & 84                                                              & 18.7                                               \\
                                                                                                                      & Otter++ (GPT-4o)                                             & 166                                                             & 37.0                                                 \\
                                                                                                                      & Otter++ (Claude-3.7-Sonnet)                                  & 174                                                             & 38.8                                               \\ \cline{2-4}
                                                                                                                      & e-Otter (GPT-4o)                                             & 167                                                             & 37.2                                               \\ 
                                                                                                                      & e-Otter (Claude-3.7-Sonnet)                                  & 179                                                             & 39.9                                               \\
                                                                                                                      & \oursystem  (GPT-4o)                                          & 231                                                             & 51.4                                               \\
                                                                                                                      & \oursystem  (Claude-3.7-Sonnet)                               & \textbf{283}                                                             & \textbf{63.0}                                                 \\ \midrule
\multirow{17}{*}{\begin{tabular}[c]{@{}l@{}}SWT-bench \\Lite (276 \\samples)\end{tabular}}                             & ZeroShot (GPT-4)                                             & 16                                                              & 5.8                                                \\
                                                                                                                      & LIBRO (GPT-4)                                                & 42                                                              & 15.2                                               \\
                                                                                                                      & AutoCodeRover (GPT-4)                                        & 25                                                              & 9.1                                                \\
                                                                                                                      & SWE-Agent+ (GPT-4)                                           & 53                                                              & 19.2                                               \\
                                                                                                                      & Otter++ (GPT-4o)                                             & 80                                                              & 29.0                                                 \\
                                                                                                                      & Issue2Test (GPT-4o-mini)                                     & 84                                                              & 30.4                                               \\
                                                                                                                      & Otter++ (Claude-3.7-Sonnet)                                  & 84                                                              & 30.4                                               \\
                                                                                                                      & AssertFlip (GPT-4o)                                     & 99                                                              & 36.0                                               \\
                                                                                                                      & Amazon Q Developer Agent* (Unknown)                          & 104                                                             & 37.7                                               \\ \cline{2-4}
                                                                                                                      & e-Otter (GPT-4o)                                             & 80                                                              & 29.0                                                 \\
                                                                                                                      & e-Otter (Claude-3.7-Sonnet)                                  & 99                                                              & 36.0                                               \\
                                                                                                                      & \oursystem  (GPT-4o)                                          & 111                                                             & 40.2                                               \\
                                                                                                                      & \oursystem  (Claude-3.7-Sonnet)                               & \textbf{145}                                                             & \textbf{52.5}                                              \\ \bottomrule
\multicolumn{4}{l}{\begin{tabular}[c]{@{}l@{}}*Number is taken from SWT-bench leaderboard.\end{tabular}}
\end{tabular}
}
\label{tbl:rq1}
\vspace{-10pt}
\end{table}

\cref{tbl:rq1} compares e-Otter and \oursystem with prior approaches on both TDD-Bench Verified and SWT-bench Lite. 
For TDD-Bench Verified and SWT-bench Lite, the best baselines report 38.8\% and 37.7\%, respectively. 
Our Claude-3.7-Sonnet based \oursystem achieves state-of-the-art performance on both benchmarks, \emph{achieving 63.0\% and 52.5\% on TDD-Bench Verified and SWT-bench Lite}, respectively, using CodePatch + Coverage as selector (see~\cref{tbl:codepatch}). 
We perform McNemar's test~\cite{mcnemar1947note} between our best test generator and all prior approaches and it significantly outperforms all the baselines on both TDD-Bench Verified and SWT-bench Lite with a 99\% confidence interval ($p<0.01$).
McNemar's test compares the performance of two systems on the same set of binary-labeled instances, highlighting statistically significant differences in their error patterns.
We also report our performance with GPT-4 and GPT-4o to demonstrate model agnosticity. 
We find that the GPT-4o-based \oursystem significantly outperforms all prior GPT-4 and GPT-4o-based solutions. 
\cref{tbl:rq1} omits some relatively weaker approaches from both benchmarks to reduce clutter.
Apart from the works listed in~\cref{tbl:rq1}, the SWT-bench Leaderboard has another entry, AEGIS, but hides it by default unless a ``script mode'' toggle is selected.
AEGIS tests are not integrated with the project's testing framework, unlike other leaderboard entries and \oursystem.
Under that setting, AEGIS reports a 47.8\% \failtopass rate with an undisclosed model.
Integrating tests with a testing framework enables future regression testing and coverage measurements.
Although not directly comparable to AEGIS, \oursystem with Claude-3.7-Sonnet outperforms AEGIS on SWT-bench Lite with a 52.5\% \failtopass rate.

While the SWT-bench paper~\cite{mundler_et_al_2024} only describes the Lite subset of their benchmark, the leaderboard recently also added a Verified subset.
SWT-bench Verified is similar to TDD-Bench Verified, as both are derived from SWE-Bench-Verified~\cite{chowdhury_et_al_2024}.
Although both benchmarks have some non-overlapping samples, systems usually perform similarly on them. 
For example, Otter++ (GPT-4o) reports 37.0\% on both TDD-Bench Verified and SWT-bench Verified. 
Given this similarity between the two ``verified'' benchmarks, we opted to use only one of them to discuss our results in this paper.

\vspace{.5em}

\findings{1}{\oursystem (Claude-3.7-Sonnet) significantly outperforms all prior approaches on TDD-Bench Verified and SWT-bench Lite, achieving 63\% and 52.5\% \failtopass rates.}

\subsection{Effectiveness of Test Repair (RQ2)}

\begin{table}[t]
\centering
\caption{Effectiveness of execution-augmented test repair}
\resizebox{1\columnwidth}{!}{%
\renewcommand{\arraystretch}{1.2}
\begin{tabular}{@{}lll|rrr|rrr@{}}
\toprule
\multicolumn{1}{@{}c}{\multirow{2}{*}{Benchmark}}                                                & \multicolumn{2}{c|}{\multirow{2}{*}{Prompt}}                                             & \multicolumn{3}{|c|}{Claude-3.7-Sonnet}                                                        & \multicolumn{3}{|c@{}}{GPT-4o}                                                                   \\ \cline{4-9}
\multicolumn{1}{@{}c}{}                                                                          & \multicolumn{2}{c|}{}                                                                    & \multicolumn{1}{|c}{Otter \failtopass} & \multicolumn{1}{c}{e-Otter \failtopass} & \multicolumn{1}{c|}{Change} & \multicolumn{1}{|c}{Otter \failtopass} & \multicolumn{1}{c}{e-Otter \failtopass} & \multicolumn{1}{c@{}}{Change} \\ \midrule
\multirow{10}{*}{\begin{tabular}[c]{@{}l@{}}TDD-Bench \\Verified (449 \\samples)\end{tabular}} & \multicolumn{2}{l|}{Planner}                                                             & 145 (32.3\%)                  & 179 (39.9\%)                    & 23.4\%                     & 141 (31.4\%)                  & 167 (37.2\%)                    & 18.4\%                     \\ \cline{2-9}
                                                                                              & \multirow{4}{*}{\begin{tabular}[c]{@{}l@{}}Context \\ Masking\end{tabular}} & full      & 125 (27.8\%)                  & 147 (32.7\%)                    & 17.6\%                     & 110 (24.5\%)                  & 135 (30.1\%)                    & 22.7\%                     \\
                                                                                              &                                                                             & testLoc   & 117 (26.1\%)                  & 130 (29.0\%)                    & 11.1\%                     & 115 (25.6\%)                  & 135 (30.1\%)                    & 17.4\%                     \\
                                                                                              &                                                                             & patchLoc  & 105 (23.4\%)                  & 134 (29.8\%)                    & 27.6\%                     & 107 (23.8\%)                  & 126 (28.1\%)                    & 17.8\%                     \\
                                                                                              &                                                                             & none      & 100 (22.3\%)                  & 127 (28.3\%)                    & 27.0\%                     & 110 (24.5\%)                  & 137 (30.5\%)                    & 24.5\%                     \\ \cline{2-9}
                                                                                              & \multirow{5}{*}{\begin{tabular}[c]{@{}l@{}}Issue \\ Morphing\end{tabular}}  & standard  & 137 (30.5\%)                  & 183 (40.8\%)                    & 33.6\%                     & 104 (23.2\%)                  & 135 (30.1\%)                    & 29.8\%                     \\
                                                                                              &                                                                             & simple    & 137 (30.5\%)                  & 182 (40.5\%)                    & 32.8\%                     & 104 (23.2\%)                  & 128 (28.5\%)                    & 23.1\%                     \\
                                                                                              &                                                                             & dropcode  & 121 (26.9\%)                  & 168 (37.4\%)                    & 38.8\%                     & 77 (17.1\%)                   & 113 (25.2\%)                    & 46.8\%                     \\
                                                                                              &                                                                             & initTests & 137 (30.5\%)                  & 166 (37.0\%)                    & 21.2\%                     & 101 (22.5\%)                  & 139 (31.0\%)                      & 37.6\%                     \\
                                                                                              &                                                                             & initPatch & 151 (33.6\%)                  & 191 (42.5\%)                    & 26.5\%                     & 119 (26.5\%)                  & 149 (33.2\%)                    & 25.2\%                     \\ \midrule
\multirow{10}{*}{\begin{tabular}[c]{@{}l@{}}SWT-bench \\Lite (276 \\samples)\end{tabular}}     & \multicolumn{2}{l|}{Planner}                                                             & 83 (30.1\%)                   & 99 (35.9\%)                     & 19.3\%                     & 70 (25.4\%)                   & 80 (29.0\%)                     & 14.3\%                     \\ \cline{2-9}
                                                                                              & \multirow{4}{*}{\begin{tabular}[c]{@{}l@{}}Context \\ Masking\end{tabular}} & full      & 67 (24.3\%)                   & 89 (32.2\%)                     & 32.8\%                     & 58 (21.0\%)                   & 71 (25.7\%)                     & 22.4\%                     \\
                                                                                              &                                                                             & testLoc   & 67 (24.3\%)                   & 82 (29.7\%)                     & 22.4\%                     & 55 (19.9\%)                   & 76 (27.5\%)                     & 38.2\%                     \\
                                                                                              &                                                                             & patchLoc  & 59 (21.4\%)                   & 72 (26.1\%)                     & 22.0\%                     & 53 (19.2\%)                   & 62 (22.5\%)                     & 17.0\%                     \\
                                                                                              &                                                                             & none      & 57 (20.7\%)                   & 68 (24.6\%)                     & 19.3\%                     & 58 (21.0\%)                   & 75 (27.2\%)                     & 29.3\%                     \\ \cline{2-9}
                                                                                              & \multirow{5}{*}{\begin{tabular}[c]{@{}l@{}}Issue \\ Morphing\end{tabular}}  & standard  & 80 (29.0\%)                   & 95 (34.4\%)                     & 18.8\%                     & 62 (22.5\%)                   & 76 (27.5\%)                     & 22.6\%                     \\
                                                                                              &                                                                             & simple    & 68 (24.6\%)                   & 88 (31.9\%)                     & 29.4\%                     & 57 (20.7\%)                   & 69 (25.0\%)                     & 21.1\%                     \\
                                                                                              &                                                                             & dropcode  & 72 (26.1\%)                   & 94 (34.1\%)                     & 30.6\%                     & 35 (12.7\%)                   & 61 (22.1\%)                     & 74.3\%                     \\
                                                                                              &                                                                             & initTests & 77 (27.9\%)                   & 99 (35.9\%)                     & 28.6\%                     & 59 (21.4\%)                   & 68 (24.6\%)                     & 15.3\%                     \\
                                                                                              &                                                                             & initPatch & 73 (26.4\%)                   & 92 (33.3\%)                     & 26.0\%                     & 68 (24.6\%)                   & 76 (27.5\%)                     & 11.8\%         \\ \bottomrule            
\end{tabular}
}
\label{tbl:rq2}
\end{table}

\cref{tbl:rq2} shows the \failtopass rate for each of the 10~tests
before~(Otter) and after~(e-Otter) execution-augmented test repair.
For example, on TDD-Bench Verified, starting from the default Planner
test, using Claude-3.7-Sonnet to repair tests improves performance
from 32.3\% to 39.9\% \failtopass rate, a relative change of~23.4\%.
While the improvements vary by test, model, and benchmark, they are
strong across the board, with relative changes ranging from a minimum
of 11.1\% to a maximum of~74.3\%.


\paragraph{Contributions of Different Components of Execution-augmented Test Repair.}

\cref{tbl:ablation-execution} shows the contribution of each component of execution-augmented test repair.
Due to cost and resource constraints, we performed this experiment with the GPT-4o model on TDD-Bench Verified only.
The row labeled e-Otter shows the performance starting with the Planner test and all repair components intact.
The remaining rows show the performance when individual components are selectively disabled.
The results show that each component contributes 7.8\% to 21.0\% to e-Otter.
To see the effectiveness of the critic, we remove the critic and fix the problem only once instead of making several attempts because the subsequent attempts depend on the critic. 
For temperature evaluation, we simply replace 0.8 with~0, i.e., greedy.
Note that the critic plays the most important role in test repair.

\begin{table}[t]
\centering
\caption{Contributions of different components of execution-augmented test repair (ablation)}
\resizebox{.7\columnwidth}{!}{%
\renewcommand{\arraystretch}{1.2}
\begin{tabular}{lrr}
\toprule  
\multicolumn{1}{c}{Approach} & \multicolumn{1}{c}{\failtopass} & \multicolumn{1}{c}{Change} \\ \midrule
e-Otter                            & 167 (37.2\%)                                                                                    & -                               \\ \midrule
without Buggy Line Selection       & 152 (33.9\%)                                                                                    & -9.0\%                               \\
without Relevant Line Extraction   & 154 (34.3\%)                                                                                    & -7.8\%                             \\
without Function Lookup              & 153 (34.1\%)                                                                                    & -8.4\%                             \\
without Critic                  & 132 (29.4\%)                                                                                    & -21.0\%                              \\
without High Temperature (0.8)     & 155 (34.5\%)                                                                                    & -7.2\%       \\ \bottomrule                     
\end{tabular}
}
\label{tbl:ablation-execution}
\end{table}

\begin{table}[t]
\centering
\caption{LLMs as critic in test repair}
\resizebox{.7\columnwidth}{!}{%
\renewcommand{\arraystretch}{1.2}
\begin{tabular}{lrrrrrr}
\toprule    
\multicolumn{1}{c}{Approach} & \multicolumn{1}{c}{TP} & \multicolumn{1}{c}{FP} & \multicolumn{1}{c}{FN} & \multicolumn{1}{c}{Precision} & \multicolumn{1}{c}{Recall} & \multicolumn{1}{c}{F-Score} \\ \midrule
Golden Test                  & 321                    & 0                      & 128                    & 1.00                          & 0.71                       & 0.83                        \\
Otter                        & 127                    & 49                     & 14                     & 0.72                          & 0.90                       & 0.80                        \\
Otter++                      & 151                    & 68                     & 15                     & 0.69                          & 0.91                       & 0.78     \\ \bottomrule                 
\end{tabular}
}
\label{tbl:llmjudge}
\end{table}

\paragraph{How good are LLMs as critics?} We just saw that critics play an important role in test repair. 
But how good are LLMs as critics? We use a GPT-4o based critic on different tests and see how it performs. 
We take GPT-4o based Otter, Otter++, and golden (i.e., developer-written) tests from TDD-Bench Verified. 
We feed the test and test log and ask the model to comment on whether the test is failing for the right 
reason mentioned in the issue description.
\cref{tbl:llmjudge} shows that GPT-4o achieves a 0.78 to~0.83 F-score as a critic, which indicates the benefit of using LLMs as critics.
Golden tests have no false positives, resulting in a precision of 1. 
However, they can have false negatives (if the model mistakenly indicates that the test is failing for the wrong reason).

\begin{figure}[t]
  \centerline{\includegraphics[width=.8\columnwidth]{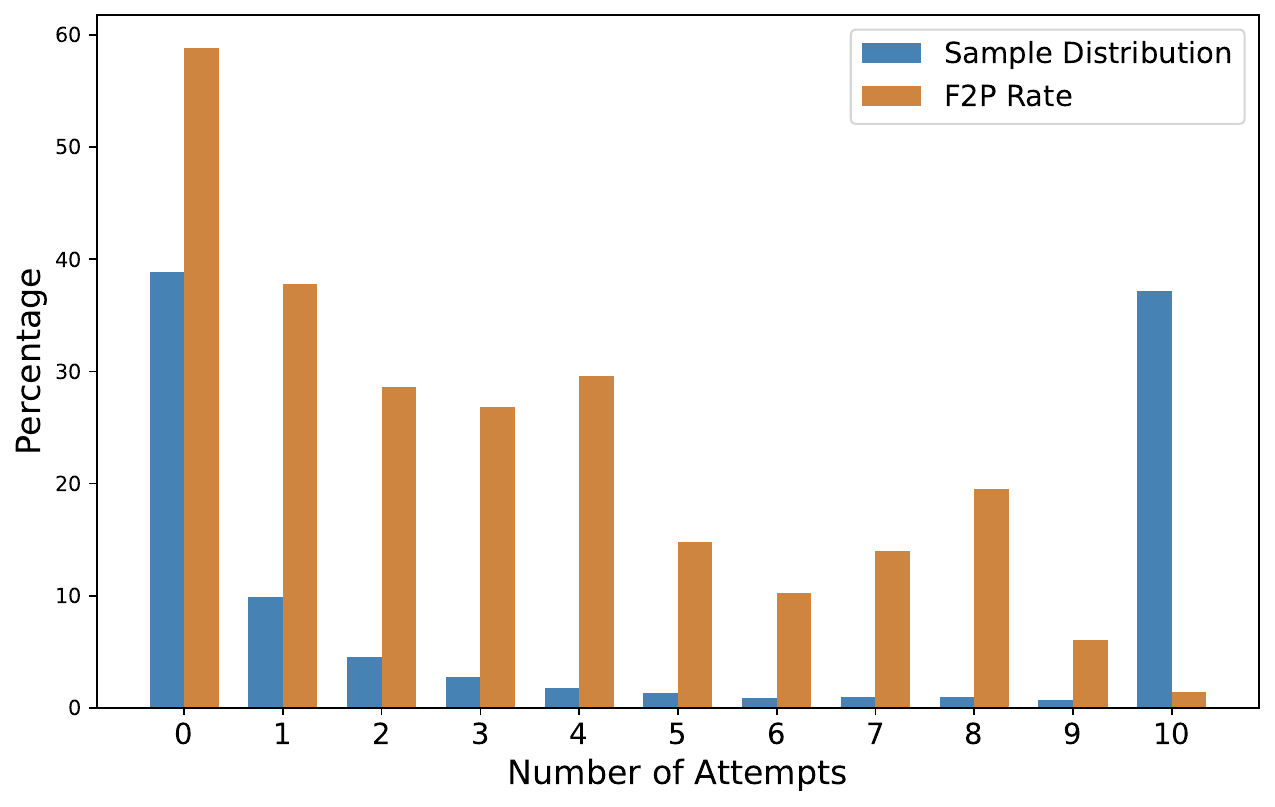}}
  \caption{\label{attempt} Number of attempts taken in test repair }
\vspace{-10pt}  
\end{figure}

\paragraph{How many attempts are required to repair the test?}
The execution-augmented test repair loop iterates for at most 10~attempts.
We measured the actual number of attempts on the 449 instances of
TDD-Bench Verified with all 5 morphs + 5 masks, i.e., 4,490 tests.
\cref{attempt} shows the highest sample distributions at 0 or 10~attempts.
There are many tests which are already \failtopass and do not need any repair. 
If the critic identifies this situation correctly, there are zero repair attempts.
For many tests, even after 10 attempts, the critic is not satisfied or is only satisfied at the $10^\textrm{th}$ attempt, and those are presented by ``10'' in the plot. 
The \failtopass rate is at least 10\% if the critic stops at 8 attempts. We can certainly go beyond 10 attempts, but after 8 attempts, the \failtopass rate drops significantly and is less than 2\% at 10~attempts.

\vspace{.5em}

\findings{2}{e-Otter significantly improves the \failtopass rate. The improvement ranges from 11.1\% to 74.3\% based on the model and benchmark under consideration.}

\subsection{Impact of Heterogeneous Prompts (RQ3)}

All \failtopass numbers we reported so far were ``@ 1'', measuring
the rate at which a single test is fail-to-pass.
Here, we instead explore \mbox{\failtopass @ N}, i.e., the rate at
which at least one among N generated tests is fail-to-pass.
That is an upper bound for how well a system can do by picking a
single test to submit from multiple generated candidate tests.
Morphs and masks yield heterogeneous prompts, which in turn increase
the diversity of candidate tests, thus hopefully increasing
\mbox{\failtopass @ N}.
We reuse 5 heterogeneous prompts including masks from
Otter++~\cite{ahmed_et_al_2025} and add 5~more prompts using morphs
for a total of 10 heterogeneous prompts.
\cref{tbl:hetero} shows that morphs increase \mbox{\failtopass @ N},
even without masks, both before~(Otter) and after~(e-Otter) test repair.
On e-Otter, the Claude-3.7-Sonnet model with \mbox{``plan + mask + morph''}
increases \mbox{\failtopass @ N} by 17.1\% and 23.8\% compared to
\mbox{``plan + mask''} on SWT-bench Lite and TDD-Bench Verified,
respectively.
We observe similar improvements with the GPT-4o model on both benchmarks.

\begin{table}[t]
\centering
\caption{Impact of heterogeneous prompts (morphs \& masks)}
\resizebox{\columnwidth}{!}{%
\renewcommand{\arraystretch}{1.2}
\begin{tabular}{@{}lll|r|rr|rr@{}}
\toprule    
\multicolumn{1}{@{}c}{Benchmark}                                                  & \multicolumn{1}{c}{Model}                                                     & \multicolumn{1}{c|}{Category} & \multicolumn{1}{|c|}{N} & \multicolumn{1}{|c}{Otter  \failtopass @ N} & \multicolumn{1}{c|}{Change} & \multicolumn{1}{|c}{e-Otter  \failtopass @ N} & \multicolumn{1}{c@{}}{Change} \\ \midrule
\multirow{6}{*}{\begin{tabular}[c]{@{}l@{}}TDD-Bench\\ Verified\end{tabular}} & \multirow{3}{*}{\begin{tabular}[c]{@{}l@{}}Claude-3.7-\\ Sonnet\end{tabular}} & plan + mask                  & 5                     & 210 (46.8\%)                      & -                         & 260 (57.9\%)                        & -                         \\
                                                                               &                                                                               & plan + morph                 & 6                     & 255 (56.8\%)                      & 21.4\%                     & 305 (67.9\%)                        & 17.3\%                     \\
                                                                               &                                                                               & plan + mask + morph          & 10                    & 270 (60.1\%)                      & 28.6\%                     & 322 (71.7\%)                        & 23.8\%                     \\ \cline{2-8}
                                                                               & \multirow{3}{*}{GPT-4o}                                                       & plan + mask                  & 5                     & 197 (43.9\%)                      & -                         & 219 (48.8\%)                        & -                         \\
                                                                               &                                                                               & plan + morph                  & 6                     & 211 (47.0\%)                      & 7.1\%                      & 240 (53.5\%)                        & 9.6\%                      \\
                                                                               &                                                                               & plan + mask + morph          & 10                    & 237 (52.8\%)                      & 20.3\%                     & 260 (57.9\%)                        & 18.7\%                     \\ \midrule
\multirow{6}{*}{\begin{tabular}[c]{@{}l@{}}SWT-bench\\ Lite\end{tabular}}    & \multirow{3}{*}{\begin{tabular}[c]{@{}l@{}}Claude-3.7-\\ Sonnet\end{tabular}} & plan + mask                  & 5                     & 113 (40.9\%)                      & -                         & 146 (52.9\%)                        & -                         \\
                                                                               &                                                                               & plan + morph                 & 6                     & 141 (51.1\%)                      & 24.8\%                     & 162 (58.7\%)                        & 11.0\%                     \\
                                                                               &                                                                               & plan + mask + morph          & 10                    & 147 (53.3\%)                      & 30.1\%                     & 171 (62.0\%)                        & 17.1\%                     \\ \cline{2-8}
                                                                               & \multirow{3}{*}{GPT-4o}                                                       & plan + mask                  & 5                     & 105 (38.0\%)                      & -                         & 119 (43.1\%)                        & -                         \\
                                                                               &                                                                               & plan + morph                 & 6                     & 114 (41.3\%)                      & 8.6\%                      & 126 (45.7\%)                        & 5.9\%                      \\
                                                                               &                                                                               & plan + mask + morph          & 10                    & 124 (44.9\%)                      & 18.1\%                     & 135 (48.9\%)                        & 13.4\%                    \\ \bottomrule
\end{tabular}
}
\label{tbl:hetero}
\end{table}

\begin{figure}[t]
  \centerline{\includegraphics[width=.9\columnwidth]{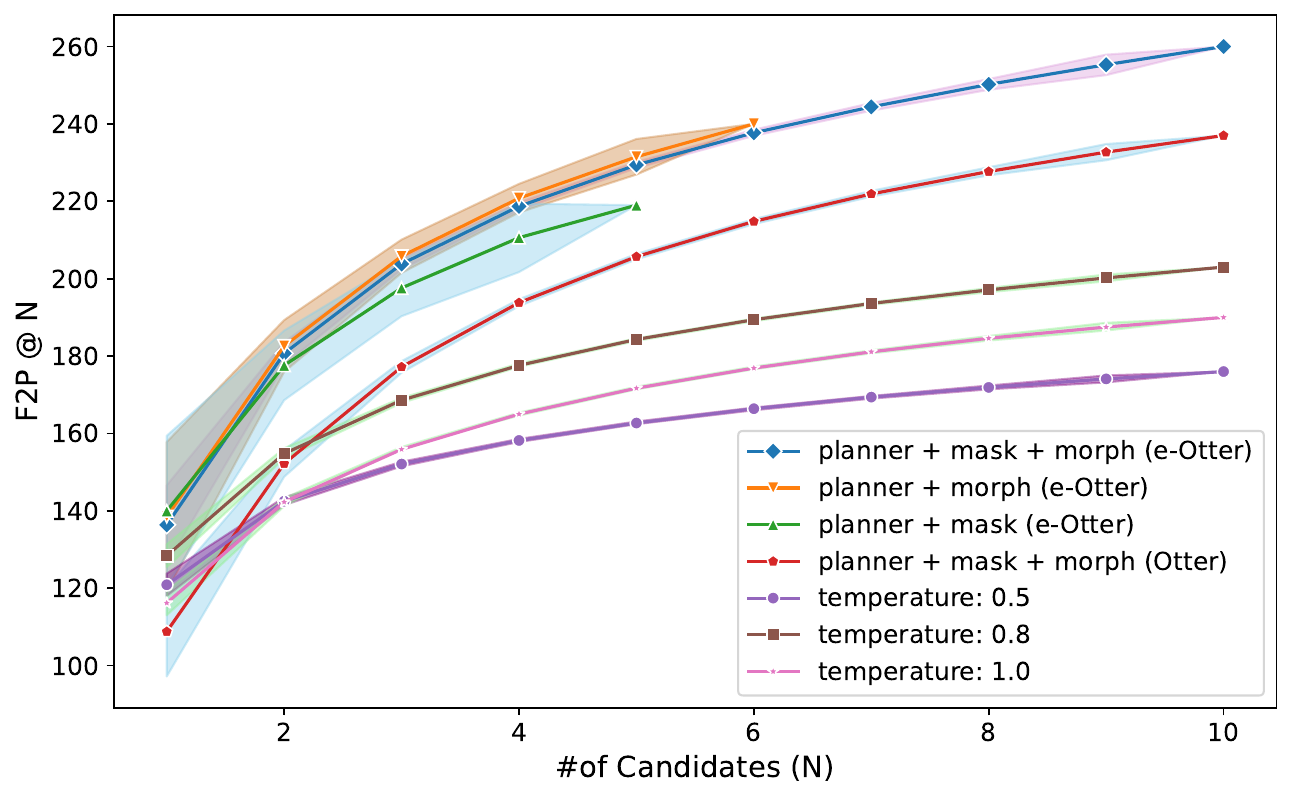}}
  \caption{\label{temperature} \failtopass @ N vs. N: a comparison of tests generated with higher temperature and heterogeneous prompting.}
\end{figure}

\paragraph{Why not use high temperature instead of heterogeneous prompting?}
Prior works~\cite{xia_et_al_2025, ehrlich_et_al_2025, chen2021evaluating} mostly depend on high temperature for multi-sampling. 
\cref{temperature} compares the results of heterogeneous prompting with the samples generated with high temperature.
We tried three distinct temperatures (0.5, 0.8, and 1.0) and find that the best results are obtained at a temperature of 0.8, which is consistent with prior research~\cite{chen2021evaluating}. 
However, our approach, \mbox{``plan + mask + morph''}, significantly outperforms samples generated with high temperature. 
In Figure~\ref{temperature}, for each $N$ (1-10), we take all possible combinations from 10 candidates and show the confidence interval in the plot.

\begin{table}[t]
\centering
\caption{Contribution of each prompt to \failtopass @ N}
\resizebox{.8\columnwidth}{!}{%
\renewcommand{\arraystretch}{1.2}
\begin{tabular}{lrrrr}
\toprule    
\multicolumn{1}{c}{\multirow{2}{*}{Approach}} & \multicolumn{2}{c}{Claude-3.7-Sonnet}                & \multicolumn{2}{c}{GPT-4o}                           \\
\multicolumn{1}{c}{}                          & \multicolumn{1}{c}{\failtopass} & \multicolumn{1}{c}{Change} & \multicolumn{1}{c}{\failtopass} & \multicolumn{1}{c}{Change} \\ \midrule
All (plan + mask + morph)                     & 322 (71.7\%)            & -                          & 260 (57.9\%)            & -                          \\
without planner                               & 316 (70.4\%)            & -1.9\%                     & 248 (55.2\%)            & -4.6\%                     \\
without full                                  & 317 (70.6\%)            & -1.6\%                     & 259 (57.7\%)            & -0.4\%                     \\
without testLoc                               & 318 (70.8\%)            & -1.2\%                     & 258 (57.5\%)            & -0.8\%                     \\
without patchLoc                              & 321 (71.5\%)            & -0.3\%                     & 259 (57.7\%)            & -0.4\%                     \\
without none                                  & 319 (71.0\%)            & -0.9\%                     & 251 (55.9\%)            & -3.5\%                     \\
without standard                              & 318 (70.8\%)            & -1.2\%                     & 256 (57.0\%)            & -1.5\%                     \\
without simple                                & 315 (70.2\%)            & -2.2\%                     & 259 (57.7\%)            & -0.4\%                     \\
without dropCode                              & 318 (70.8\%)            & -1.2\%                     & 253 (56.3\%)            & -2.7\%                     \\
without initTest                              & 317 (70.6\%)            & -1.6\%                     & 254 (56.6\%)            & -2.3\%                     \\
without initPatch                             & 312 (69.5\%)            & -3.1\%                     & 256 (57.0\%)            & -1.5\%         \\ \bottomrule            
\end{tabular}
}
\label{tbl:promptablation}
\end{table}

\paragraph{Impact of each prompt in \failtopass @ N (ablation)}
\cref{tbl:promptablation} shows the impact of each prompt for the Claude-3.7-Sonnet and GPT-4o models on TDD-Bench Verified. 
Each prompt contributes between 0.3\% and 4.6\% to \failtopass @ N.
Even though some prompts do not add many new instances, 
increasing \failtopass @ N is still beneficial, as it gives the test selector a larger pool of valid candidates to choose from, increasing the chances of selecting the most suitable ones.

Although we report most results using two closed-source models, Claude-3.7-Sonnet and GPT-4o, 
we also investigate the effectiveness of execution-augmented test repair and heterogeneous prompting on a relatively weaker open-source model, Mistral-large~(123 billion parameters). 
The \failtopass rate increases by 8.3\% to 37\% by execution-augmented test repair, and \failtopass @ N increases by 11\% with heterogeneous prompting (plan + mask + morph).
This indicates that our approaches are also applicable to open-source models.


Multiple morphs can independently benefit a single issue on TDD-bench-Verified. 
Using the base e-Otter setup, applying an additional morph increases \failtopass @ 2 
relative to \failtopass @ 1 by 29\%–32\% for Claude and 11\%–21\% for GPT-4o.


\vspace{.5em}

\findings{3}{The samples with \mbox{``plan + mask + morph''} increase \mbox{\failtopass @ N} by more than 13\% compared to \mbox{``plan + mask''}, regardless of the model and benchmark.}

\subsection{Effectiveness of Code Patches (RQ4)}

\begin{table}[t]
\centering
\caption{Test selection using code patches}
\resizebox{.9\columnwidth}{!}{%
\renewcommand{\arraystretch}{1.2}
\begin{tabular}{llllr}
\toprule    
\multicolumn{1}{c}{Benchmark}       & \multicolumn{1}{c}{Model}         & \multicolumn{1}{c}{Pre-filter} & \multicolumn{1}{c}{Selection} & \multicolumn{1}{c}{\failtopass} \\ \midrule
\multirow{8}{*}{\begin{tabular}[c]{@{}l@{}}TDD-Bench\\ Verified\end{tabular}} & \multirow{4}{*}{Clade-3.7-Sonnet} & None                           & Random              & 200.5 (44.7\%)          \\
                                    &                                   & None                           & Coverage                      & 213.0 (47.4\%)          \\
                                    &                                   & CodePatch                      & Random            & 277.6 (61.8\%)          \\
                                    &                                   & CodePatch                      & Coverage                      & 283.0 (63.0\%)          \\ \cline{2-5}
                                    & \multirow{4}{*}{GPT-4o}           & None                           & Random              & 161.6 (36.0\%)          \\
                                    &                                   & None                           & Coverage                      & 182.0 (40.5\%)          \\
                                    &                                   & CodePatch                      & Random               & 227.0 (50.6\%)          \\
                                    &                                   & CodePatch                      & Coverage                      & 231.0 (51.4\%)          \\ \midrule
\multirow{8}{*}{\begin{tabular}[c]{@{}l@{}}SWT-bench\\ Verified\end{tabular}}     & \multirow{4}{*}{Clade-3.7-Sonnet} & None                           & Random              & 100.9 (36.5\%)          \\
                                    &                                   & None                           & Coverage                      & 111.0 (40.2\%)          \\
                                    &                                   & CodePatch                      & Random              & 140.6 (50.9\%)          \\
                                    &                                   & CodePatch                      & Coverage                      & 145.0 (52.5\%)          \\ \cline{2-5}
                                    & \multirow{4}{*}{GPT-4o}           & None                           & Random            & 83.4 (30.2\%)           \\
                                    &                                   & None                           & Coverage                      & 88.0 (31.9\%)           \\
                                    &                                   & CodePatch                      & Random               & 108.4 (39.3\%)          \\
                                    &                                   & CodePatch                      & Coverage                      & 111.0 (40.2\%)     \\ \bottomrule     
\end{tabular}
}
\label{tbl:codepatch}
\end{table}

This section examines the effectiveness of test selection using
execution feedback from generated code patches to pre-filter a test
patch, as well as using coverage as a final selection tie-breaker.
The baseline for the tie-breaker is random selection, where we repeat
the experiment 10 times and report the average \failtopass rate.
\cref{tbl:codepatch} shows the results for both models on both benchmarks. 
Without code patches as a pre-filter, coverage alone improves performance by 6.2\% (from 200.5 to 213) for the Claude model on TDD-Bench Verified. 
With generated code patches as a pre-filter, coverage still provides some benefit, but the improvement is relatively low, at 1.9\% (from 277.6 to 283).
In contrast, code patches have a significant impact on performance. With coverage as a tie-breaker, the \failtopass rate increases by 32.9\% (from 213 to 283) because of pre-filtering.
Other model-benchmark combinations show similar results.

\vspace{.5em}

\findings{4}{Test selection benefits from both the use of generated code patches as a pre-filter and coverage as a tie-breaker among multiple tests within a group. 
Generated code patches play the most important role in test selection.}

\subsection{Validating SWE-patches (RQ5)}

\begin{table}[t]
\centering
\caption{Validating SWE-patches with our generated tests}
\resizebox{.9\columnwidth}{!}{%
\renewcommand{\arraystretch}{1.2}
\begin{tabular}{rlrrrrr}
\toprule  
\multicolumn{1}{c}{\multirow{2}{*}{Rank}} & \multicolumn{1}{c}{\multirow{2}{*}{System}} & \multicolumn{1}{c}{\multirow{2}{12mm}{Old Precision}} & \multicolumn{2}{c}{Claude-3.7-Sonnet}                          & \multicolumn{2}{c}{GPT-4o}                                     \\ \cline{4-7}
\multicolumn{1}{c}{}                      & \multicolumn{1}{c}{}                        & \multicolumn{1}{c}{}                               & \multicolumn{1}{c}{Recall} & \multicolumn{1}{c}{New Precision} & \multicolumn{1}{c}{Recall} & \multicolumn{1}{c}{New Precision} \\ \midrule
1                                         & tools\_claude-4-opus                        & 74.2\%                                             & 72.6\%                     & 78.8\%                            & 57.7\%                     & 81.5\%                            \\
2                                         & tools\_claude-4-sonnet                      & 73.5\%                                             & 73.3\%                     & 78.1\%                            & 59.0\%                     & 79.6\%                            \\
9                                         & sweagent\_claude-4-sonnet                   & 68.4\%                                             & 69.9\%                     & 74.5\%                            & 56.6\%                     & 74.8\%                            \\
10                                        & aime\_coder                                 & 67.9\%                                             & 58.8\%                     & 78.4\%                            & 47.9\%                     & 82.3\%                            \\
11                                        & openhands                                   & 66.1\%                                             & 72.8\%                     & 73.4\%                            & 58.8\%                     & 75.4\%                     \\ \bottomrule      
\end{tabular}
}
\label{tbl:rq5}
\end{table}

The tests generated by our approach can be used for discarding bad code patches and increasing the precision of solutions 
proposed by different systems from the SWE-bench Verified leaderboard. 
Ahmed et al.~\cite{ahmed_et_al_2025} took the top 22 systems from the leaderboard and ran the five tests generated by Otter++. 
They filtered out a code patch if all the tests failed on it.
This achieved a precision of 65\% to 92\% while maintaining a decent recall of 30\%-41\%. 
M{\"u}ndler et al.~\cite{mundler_et_al_2024} performed a similar experiment with one system. 
Note that neither of these two prior approaches could achieve a \failtopass rate beyond 50\% even with 5 samples. 
Therefore, the recall suffers significantly. 
In contrast, \oursystem achieves a 71.7\% \mbox{\failtopass @ 10} rate, which could significantly benefit the recall.
To showcase the efficacy of our tests in conjunction with top SWE agents, we pick the top SWE agents from the leaderboard where the generated code patches were available. 
Unfortunately, due to a recent change in the SWE-Bench submission process, we could not retrieve them for some systems.
\cref{tbl:rq5} shows the old precision, recall, and new precision on SWE-Bench-Verified using both Claude-3.7-Sonnet and GPT-4o models.  
We achieve 47.9\% to 72.8\% recall, regardless of the model, while substantially improving precision.

\vspace{.5em}

\findings{5}{We achieve 47.9\% to 72.8\% recall, regardless of the model, while increasing the precision of the SWE-patches.}

\subsection{Performance on SWE-rebench (RQ6)}
To show that our proposed approach is effective for non-contamina\-ted models, 
we extend our evaluation on SWE-rebench with the Claude-3.7-Sonnet model. 
The training data cut-off date for Claude-3.7-Sonnet is November 2024~\cite{claude-cutoff}. 
To address data contamination, we consider instances from SWE-rebench~\cite{badertdinov2025swe} with a creation date in 2025 and 
check whether the developer-written golden tests go from fail-to-pass to select the instances. 
This experiment uses 267~instances. 
Table~\ref{tbl:rebenchrq1} shows that execution-augmented test repair increases the fail-to-pass rate from 21\% to 72\% for all prompt variants. 
Table~\ref{tbl:rebenchrq2} also shows that heterogeneous prompting increases the fail-to-pass @ N by more than 32\%, 
which is consistent with our findings on other benchmarks.

The last step of our pipeline, test selection, uses surrogate patches.
We use leaderboard patches shared by the SWE-rebench authors instead of Agentless to evaluate this step. 
The average resolve rate of these patches is less than 40\%, and these patches are also free from contamination.
Table~\ref{tbl:rebench3} shows that \oursystem achieved a 42.7\% \failtopass rate on SWE-rebench, which is 75\% 
higher than Otter.
The absolute numbers on the non-contamina\-ted SWE-rebench are notably
lower than on TDD-Bench or SWT-bench, and contamination might be a
major factor behind that.
Another factor is that the samples from SWE-rebench come from 152
projects, most of which are less popular than the projects in SWE-bench.
Popular projects are relatively older, larger, and may be
more likely to have well-written issue descriptions.
Despite lower absolute numbers, our approach shows a great relative
improvement, demonstrating that our two primary contributions also
work for contamination-free benchmarks like SWE-rebench.

\begin{table}[t]
\centering
\caption{Effectiveness of execution feedback on SWE-rebench}
\resizebox{.7\columnwidth}{!}{%
\renewcommand{\arraystretch}{1.2}
\begin{tabular}{ll|rr|rr|r}
\toprule  
\multicolumn{2}{c}{\multirow{2}{*}{Prompt}}  & \multicolumn{2}{|c|}{Otter}                                  & \multicolumn{2}{c|}{e-Otter}                                & \multicolumn{1}{c}{\multirow{2}{*}{Change}} \\
\multicolumn{2}{c}{}                         & \multicolumn{1}{|c}{\# of  F-P} & \multicolumn{1}{c|}{in \%} & \multicolumn{1}{c}{\# of  F-P} & \multicolumn{1}{c|}{in \%} & \multicolumn{1}{c}{}                        \\ \midrule
\multicolumn{2}{l|}{Planner}                  & 65                             & 24.3                      & 86                             & 32.2                      & 32.3\%                                        \\ \midrule
\multirow{4}{*}{Context Masking} & full      & 56                             & 21.0                      & 71                             & 26.6                      & 26.8\%                                        \\
                                 & testLoc   & 49                             & 18.4                      & 67                             & 25.1                      & 36.7\%                                        \\
                                 & patchLoc  & 41                             & 15.4                      & 60                             & 22.5                      & 46.3\%                                        \\
                                 & none      & 25                             & 9.4                       & 43                             & 16.1                      & 72.0\%                                          \\ \midrule
\multirow{5}{*}{Issue Morphing}  & standard  & 60                             & 22.5                      & 73                             & 27.3                      & 21.7\%                                        \\
                                 & simple    & 45                             & 16.9                      & 68                             & 25.5                      & 51.1\%                                        \\
                                 & dropcode  & 49                             & 18.4                      & 75                             & 28.1                      & 53.1\%                                        \\
                                 & initTests & 50                             & 18.7                      & 72                             & 27.0                        & 44.0\%                                          \\ 
                                 & initPatch & 56                             & 21.0                        & 70                             & 26.2                      & 25.0\%     \\ \bottomrule                                    
\end{tabular}
}
\label{tbl:rebenchrq1}
\vspace{-5pt}
\end{table}

\begin{table}[t]
\centering
\caption{Impact of heterogeneous prompts on SWE-rebench}
\resizebox{\columnwidth}{!}{%
\renewcommand{\arraystretch}{1.2}
\begin{tabular}{l|r|rrr|rrr}
\toprule  
\multicolumn{1}{c}{\multirow{2}{*}{Category}} & \multicolumn{1}{|c|}{\multirow{2}{*}{N}} & \multicolumn{3}{|c|}{Otter}                                                                     & \multicolumn{3}{c}{e-Otter}                                                                   \\
\multicolumn{1}{c}{}                          & \multicolumn{1}{|c|}{}                   & \multicolumn{1}{c}{\failtopass @ N} & \multicolumn{1}{c}{In \%} & \multicolumn{1}{c|}{Change} & \multicolumn{1}{c}{\failtopass @ N} & \multicolumn{1}{c}{In \%} & \multicolumn{1}{c}{Change} \\ \midrule
plan + mask                                   & 5                                      & 97                                   & 36.3                      & NA                         & 119                                  & 44.6                      & NA                         \\
plan + morph                                  & 6                                      & 115                                  & 43.1                      & 18.6\%                       & 146                                  & 54.7                      & 22.7\%                       \\
plan + mask + morph                           & 10                                     & 129                                  & 48.3                      & 33.0\%                         & 158                                  & 59.2                      & 32.8\%                      \\ \bottomrule
\end{tabular}
}
\label{tbl:rebenchrq2}
\vspace{-5pt}
\end{table}

\begin{table}[t]
\centering
\caption{Performance of \oursystem on SWE-rebench}
\resizebox{.65\columnwidth}{!}{%
\renewcommand{\arraystretch}{1.2}
\begin{tabular}{lrrr}
\toprule  
\multicolumn{1}{c}{Approach} & \multicolumn{1}{c}{\# of \failtopass} & \multicolumn{1}{c}{in \%} & \multicolumn{1}{c}{Change from Otter} \\ \midrule
Otter                        & 65                                 & 24.3                      & NA                                    \\
Otter++                      & 87                                 & 32.6                      & 33.8\%                                  \\ \midrule
e-Otter                      & 86                                 & 32.2                      & 32.3\%                                  \\
e-Otter++                    & 114                                & 42.7                      & 75.4\%          \\ \bottomrule                       
\end{tabular}
}
\label{tbl:rebench3}
\vspace{-10pt}
\end{table}

\findings{6}{\oursystem is effective regardless of the data contamination issue. 
It improves the \failtopass rate by a relative 75\% compared to Otter on SWE-rebench.}

%% file: discussion.tex
\section{Discussion}
\label{sec:discusion}

This section discusses model contamination, test coverage, and a detailed analysis of the tests.

\vspace{3pt}

\noindent\textbf{Contamination through Issue Morphs.} Test generation from issues could potentially suffer from model memorization and contamination since the model might have seen the data during the training phase. 
Both prior works by Ahmed et al.~\cite{ahmed_et_al_2025} and M{\"u}ndler et al.~\cite{mundler_et_al_2024} addressed this issue and show that the model-generated tests are very different 
from developer-written tests and the impact of contamination is minimal. 
However, \oursystem might
potentially introduce new contamination.
Specifically, issue morphs might introduce contamination while
rewriting the issue.
To investigate this, 
we manually validated the samples before execution-augmented test repair. 
We took 50 samples (25 from each model) where the original planner prompt was not producing \failtopass tests but after issue morph the tests became \failtopass. 
We had 4 raters with programming experience ranging from 15-30 years. We presented the raters with four pieces of information: (i)~original issue description,
(ii)~modified issue, (iii)~non \failtopass version of the test, (iv)~\failtopass version of the test. We asked the raters the following three questions.

\begin{itemize}[leftmargin=2em]

\item Is the modified issue adding new information that might influence the model's decision?
\item Is the modified issue removing any information that might influence the model's decision?
\item In cases of addition, does this rewrite cause contamination that might help the model generate fail-to-pass tests? The rater can decide on this by comparing the two versions of the tests.

\end{itemize}

The raters initially rated 5 samples and had a discussion on how to rate the remaining samples. 
After that, each rater individually rated the samples. Since we have four raters, 
it is infeasible to do majority voting on the rated instances. 
Instead, we adopted a stricter metric to assign the final label. If at least two raters think there is some addition, removal, or contamination, we counted it as ``1'', otherwise ``0''.
\cref{tbl:contamination} shows the inter-rater agreement and outcome of the study. 
We use Fleiss' Kappa~\cite{fleiss1971measuring} to compute inter-rater agreement, and the raters show fair to substantial agreement\footnote{Kappa < 0: Poor, 0.01–0.20: Slight, 0.21–0.40: Fair, 0.41–0.60: Moderate, 0.61–0.80: Substantial, 0.81–1.00: Almost perfect agreement.}.
Note that for contamination, we observe a very low kappa value because all raters mostly choose the same category. This phenomenon is known as the kappa paradox~\cite{feinstein1990high}.  
Overall, 40\% of modified issues have some information addition and 68\% of issues have some information removal. 
Claude-3.7-Sonnet makes more additions and removals compared to GPT-4o. 
However, none of the raters think there is substantial contamination due to issue morphs. Only for 1 (2\%) sample out of 
50 samples, more than one rater think the rewriting might have caused some contamination. 
Therefore, we can infer that issue morphs do not contaminate the issue. 

\begin{table}[t]
\centering
\caption{Manual validation to investigate model contamination or memorization}
\resizebox{.7\columnwidth}{!}{%
\renewcommand{\arraystretch}{1.2}
\begin{tabular}{llrrlr}
\toprule  
\multicolumn{1}{c}{\multirow{2}{*}{Category}} & \multicolumn{1}{c}{\multirow{2}{*}{Model}} & \multicolumn{1}{c}{\multirow{2}{*}{N}} & \multicolumn{2}{c}{Fleiss' Kappa}                                                                                                                                      & \multicolumn{1}{c}{\multirow{2}{*}{Count}} \\  \cline{4-5}
\multicolumn{1}{c}{}                          & \multicolumn{1}{c}{}                       & \multicolumn{1}{c}{}                   & \multicolumn{1}{c}{Value}                                                     & \multicolumn{1}{c}{Interpretation}                                                     & \multicolumn{1}{c}{}                       \\ \midrule
\multirow{3}{*}{Addition}                     & Claude-3.7-Sonnet                          & 25                                     & 0.33                                                                          & Fair                                                                                   & 16 (64\%)                                  \\
                                              & GPT-4o                                     & 25                                     & 0.37                                                                          & Fair                                                                                   & 4 (16\%)                                   \\
                                              & Both                                       & 50                                     & 0.44                                                                          & Moderate                                                                               & 20 (40\%)                                  \\ \midrule
\multirow{3}{*}{Removal}                      & Claude-3.7-Sonnet                          & 25                                     & 0.57                                                                          & Moderate                                                                               & 19 (76\%)                                  \\
                                              & GPT-4o                                     & 25                                     & 0.63                                                                          & Substantial                                                                            & 15 (30\%)                                  \\
                                              & Both                                       & 50                                     & 0.60                                                                          & Moderate                                                                               & 34 (68\%)                                  \\ \midrule
\multirow{3}{*}{Contamination}                & Claude-3.7-Sonnet                          & 25                                     & \multicolumn{2}{l}{\multirow{3}{*}{\begin{tabular}[c]{@{}l@{}}Kappa value is not meaningful \\ because all raters mostly choose \\ the same category.\end{tabular}}} & 0 (0\%)                                    \\
                                              & GPT-4o                                     & 25                                     & \multicolumn{2}{l}{}                                                                                                                                                   & 1 (4\%)                                    \\
                                              & Both                                       & 50                                     & \multicolumn{2}{l}{}                                                                                                                                                   & 1 (2\%)                                
\\ \bottomrule
                                            \end{tabular} 
}
\label{tbl:contamination}
\end{table}

\begin{figure}[t]
  \centerline{\includegraphics[width=.80\columnwidth]{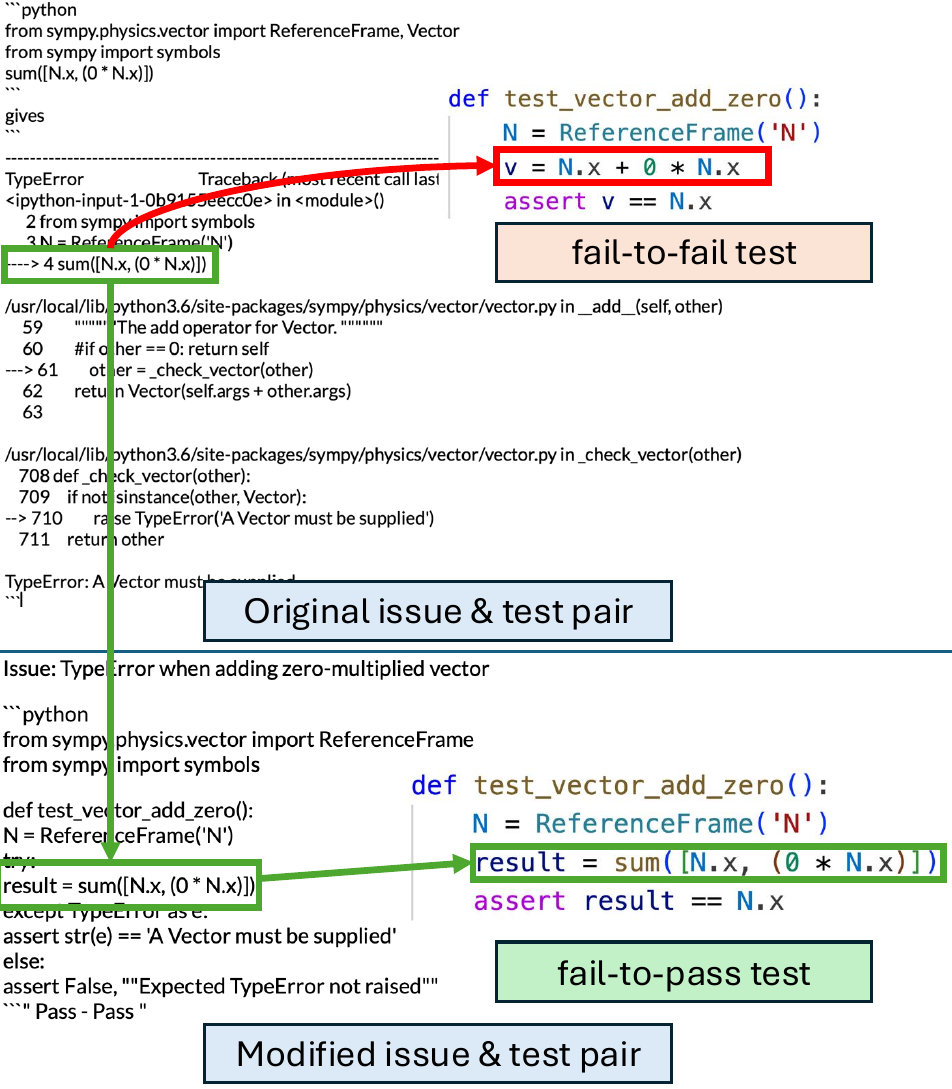}}
  \caption{\label{example-contamination} Example initTest (sympy\_\_sympy-14711).}
\end{figure}

\cref{example-contamination} shows the example that was marked by
three out of four raters as possible contamination.
When we took a closer look at it, we became uncertain about it.
In this particular case, the LLM was asked to generate an initial test with the issue description. The fixed line is ``result = sum([N.x, (0 * N.x)])''. 
This line is present in the modified issue description and that might have inspired the raters to mark it as a suspicious test. 
However, that same line was also present in the original issue description. 
The original test considered that line buggy and opted for a different version of this line. 
Seeing the fixed line in the proposed test definitely helps the model to use the right version. 
But whether this case constitutes actual contamination is unclear since that line is present in the original issue description.

\begin{figure}[t]
  \centerline{\includegraphics[width=.8\columnwidth]{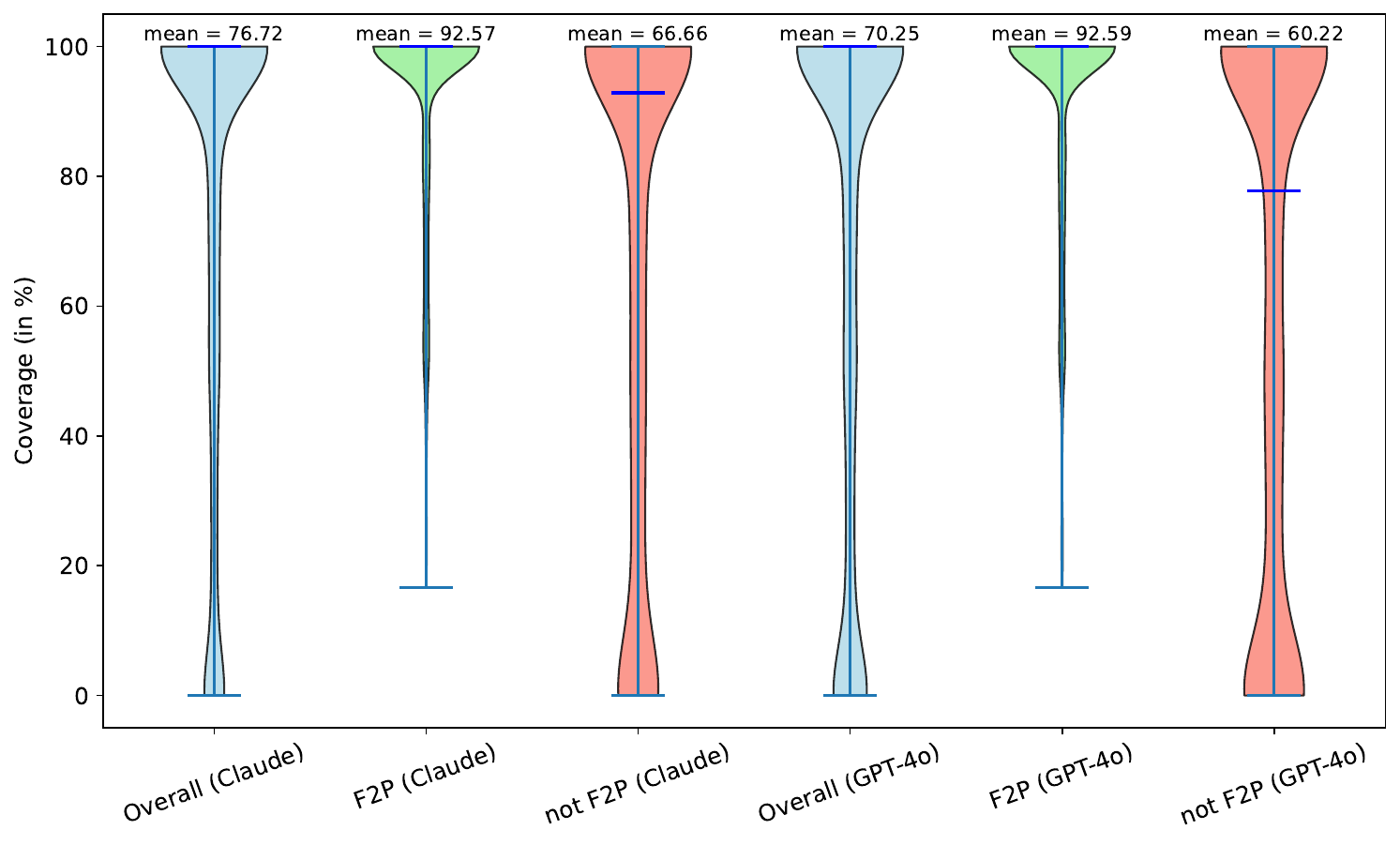}}
  \caption{\label{coverage} Coverage of the generated tests}
\end{figure}

\vspace{3pt}
\noindent\textbf{Coverage of the Generated Tests.} Coverage is an important metric to measure the adequacy of tests. 
We measure how well a generated test covers the developer-written
golden code patch (note that this is not available during test
generation).
To do that, we find the deleted lines in~$c_\textrm{old}$ and added
lines in~$c_\textrm{new}$ and see whether the test covers those lines.
We take the total number of covered lines and divide them by the total number of deleted and added lines to compute the coverage. 
\cref{coverage} shows that \failtopass tests have higher coverage compared to other tests.  
These findings are consistent with prior works~\cite{ahmed_et_al_2025, mundler_et_al_2024}
and motivate using coverage as a tie-breaker for test selection
(of course, test selection uses coverage on generated code patches,
not golden ones).

\vspace{3pt}

\noindent\textbf{Detailed Analysis of the Tests.} Table\ref{tbl:analysis} presents an analysis of the e-Otter-generated tests using GPT-4o from
different perspectives. 
We did not see much impact
of focal localization on the performance, with 37.8\% and
34.2\% for correct and incorrect localization, respectively.
Test localization has a significant impact on the performance
(42.6\% vs. 24.2\% for correct and incorrect localization). 
We see that the fail-to-pass rate is
higher when modify an existing test (40.2\%) than when writing a
new test (36.1\%), which is not surprising because the model can have
better context from the existing test. It is
expected that the test will fail on the old codebase. However,
we found 43 samples in e-Otter where the test passed on $c_\textrm{old}$. 
Also, our analysis shows that tests with assertion failures have a higher success rate (50.4\% \failtopass
rate) compared to other groups.

\begin{table}[t]
\centering
\caption{Analysis of the e-Otter-generated tests using GPT-4o model from different perspectives.}
\resizebox{.7\columnwidth}{!}{%
\renewcommand{\arraystretch}{1.2}
\begin{tabular}{llrrr}
\toprule    
\multicolumn{1}{c}{Perspective}       & \multicolumn{1}{c}{Category} & \multicolumn{1}{c}{\#of Sample} & \multicolumn{1}{c}{\failtopass} & \multicolumn{1}{c}{In (\%)} \\ \midrule

\multirow{2}{*}{Focal Localization}   & Correct                      & 370                             & 140                     & 37.8                        \\
                                      & Wrong                        & 79                              & 27                      & 34.2                        \\ \midrule
\multirow{2}{*}{Test Localization}    & Correct                      & 317                             & 135                     & 42.6                        \\
                                      & Wrong                        & 132                             & 32                      & 24.2          \\  \midrule
\multirow{2}{*}{Type of Test}         & Modified                     & 122                             & 49                      & 40.2                        \\
                                      & New                          & 327                             & 118                     & 36.1                        \\ \midrule
\multirow{4}{*}{Test on old Codebase} & Pass                         & 43                              & 0                       & 0                           \\
                                      & AssertionFail                & 224                             & 113                     & 50.4                        \\
                                      & Fail                         & 94                              & 46                      & 48.9                        \\
                                      & Error                        & 88                              & 8                       & 9.1                         \\

                                      \bottomrule              
\end{tabular} 
}
\label{tbl:analysis}
\end{table}

\vspace{3pt}

\noindent\textbf{Cost and Time.} Our system, \oursystem, uses both sequential inference scaling~(via
test repair) and parallel inference scaling (via heterogeneous
prompting). 
Furthermore, we also use Agentless code patches for test selection, which requires several LLM calls. 
We did not generate the code patches for Agentless ourselves; instead, we relied on code patches shared by the authors for both models. 
However, based on their reported numbers and our experiments, we can estimate the cost for each sample. 
For the Claude and GPT-4o models, our estimated cost is \$2.75 and \$1.80 per instance, respectively.
Note that prior test generation approaches do not use code patches for test selection, which adds more cost. 
However, it is likely that a reproduction test generation approach will be paired with a SWE-patch generation approach. 
In that case, the additional cost of generating code patches will be amortized.
Additionally, excluding Agentless, e-Otter++ costs \$1.70 (\$0.17/candidate) for Claude and \$1.10 (\$0.11/candidate) for GPT-4o. 
Figure~\ref{temperature} shows how the \failtopass @ N changes with the number of candidates. 
For example, 6 samples yield 230 \failtopass tests and generating 6 samples costs \$0.66.
It is hard to report the end-to-end time because several steps of our process can be parallelized. 
We did most of our experiments sequentially due to shared resources. 
Docker setup takes 5-10 minutes based on the repository.
Other steps (i.e., test variant generation, agentless generation, ranker) can take up to 2–3 minutes, some of which are parallelizable.

%% file: threats.tex
\section{Threats to Validity}

One limitation of our work is that it only experiments with Python. 
As a result, the findings may not generalize to other programming languages. 
However, prior works~\cite{jimenez_et_al_2024, xia_et_al_2025, zhang_et_al_2024} with the same limitation have still been impactful and contributed significantly to the field.
One of the key concerns with SWE-bench related work is the contamination or memorization problem of the model~\cite{badertdinov2025swe,liang2025swe}. 
To address this problem, we evaluated \oursystem on SWE-rebench~\cite{badertdinov2025swe} and observed similar impact 
from execution-augmented test repair and heterogeneous prompting.
(We did not use REPOCOD~\cite{liang_et_al_2025}, since it is not contamination-free, nor LiveCodeBench~\cite{jain_et_al_2025-livecodebench}, since it is not repository-level.)
Another limitation
of \oursystem is that it considers only one test file and generates
only one block of code.
In real-world projects, test code can
be spread across multiple files or blocks of code. Despite
that, \oursystem exceeds the state-of-the-art performance, so
we leave further improvements to future work.

It is hard to ensure the correctness of the generated tests. 
TDD-bench-Verified and SWT-bench use coverage in conjunction with the \failtopass property to be considered a success. 
Coverage is a primary metric for test quality. 
These benchmarks allow us to apply the golden patch (written by real developers and likely to be correct), 
run the test, and check both how well it covers the edited lines and whether it goes from fail to pass. 
While this is no correctness guarantee, it is a strong proxy for correctness and followed by prior works~\cite{mundler_et_al_2024, ahmed_et_al_2025}.
Besides, using imperfect patches or tests comes with certain risks, 
but prior work~\cite{chen_et_al_2023} has shown that dual execution agreement 
can help discard samples with potential errors.
To further mitigate risk, one could use separate systems for surrogate patches and 
the proposed final patch (of course, with increased cost).

%% file: relatedwork.tex
\section{Related Work}

There has been a flurry of recent papers about generating reproduction
tests from issues, which is the same problem we tackle in our paper.
Libro generates tests for Java bug reports~\cite{kang_yoon_yoo_2023}.
It repairs tests based on compiler feedback, not execution feedback
like we do, and generates several tests using high LLM decoding
temperature, not morphs and masks like we do.
Plein et al.\ also explore the feasibility of generating tests for
Java bug reports, albeit without test repair~\cite{plein_et_al_2024}.
M{\"u}ndler et al.~\cite{mundler_et_al_2024} introduce SWE-Agent+, a
Python reproduction test generation agent derived from
the SWE-Agent~\cite{yang_et_al_2024} code patch generator.
While SWE-Agent+ has access to tools that can give it execution
feedback, it leverages such feedback less effectively than \oursystem.
The same holds for EvoCoder~\cite{lin_et_al_2024}, also derived from
SWE-Agent, and for USEAgent~\cite{applis_et_al_2025}, derived from
AutoCodeRover~\cite{zhang_et_al_2024}.
AEGIS is a novel multi-agent system for reproduction test
generation~\cite{wang_et_al_2024}.
It repairs tests based on execution feedback from $c_\textrm{old}$,
but unlike our work, does not augment the test repair prompt.
Also, it standardizes issue descriptions, but unlike our work, does
not use morphs or masks to increase test candidate pool diversity.
Otter++~\cite{ahmed_et_al_2025} is an LLM-based workflow for
reproduction test generation that we use as a component in
\oursystem, which extends it in three important ways:
(i)~augmenting the test repair prompt;
(ii)~using morphs (Otter++ only uses masks); and
(iii)~leveraging execution feedback from generated candidate code patches.
Our experiments show that these novel contributions greatly improve
the \failtopass rate.
BRT Agent is a new agent for generating reproduction
tests~\cite{cheng_et_al_2025}, and Issue2Test is a new non-agentic
workflow for the same task~\cite{nashid_et_al_2025};
neither augments test repair prompts, uses masks or morphs, or
leverages candidate code patches.

Unlike \oursystem and the above papers on using generated code patches to
help pick or improve reproduction tests, there is also work on the opposite:
using reproduction tests to help pick or improve code patches.
Papers on using tests to help pick (but not improve) code patches
include Smith et al.'s 2015 paper~\cite{smith_et_al_2015},
CodeT~\cite{chen_et_al_2023}, Agentless~\cite{xia_et_al_2025},
and R2E-Gym~\cite{jain_et_al_2025}.
Similarly, several papers use tests to help improve (but not pick) code
patches~\cite{zhang_et_al_2024,yang_et_al_2024,chen_et_al_2024,arora_et_al_2024,ruan_zhang_roychoudhury_2024,mathews_nagappan_2024}.
Some papers even do both, using tests to help pick \emph{and} improve code
patches:
CodeMonkeys~\cite{ehrlich_et_al_2025} runs several parallel workflows,
each of which iteratively refines a \mbox{$\langle c_\mathrm{new},y\rangle$}
pair; and
PatchPilot~\cite{li_et_al_2025} uses reproduction tests for bug
localization before code patch generation, and later
uses them again to pick among multiple candidate code patches.
However, while most of the work described in this paragraph uses generated
tests, test generation is not their main focus:
their reproduction test generation components are more basic, and they
lack empirical evaluation of the \failtopass rate of generated tests.
That said, they demonstrate a widespread need for our work.
Besides, Chen et al.~\cite{chen2025old} show that, together with regression tests, 
bug reproduction tests can improve the performance of issue resolution systems.


%% file: conclusion.tex
\section{Conclusion}

This paper introduces \oursystem, a system for generating reproduction
tests from SWE issues.
It works by first generating a set of candidate tests, with novel
techniques for increasing diversity by morphing issue descriptions and
masking parts of the test generation context.
Next, it iteratively repairs each test using feedback from executing
it on the old code; what is novel in this step is that it augments the
test repair prompt with new context based on the execution feedback.
Finally, it selects a single final reproduction test; since the new
code that fixes the issue does not yet exist, it uses execution
feedback from multiple generated candidate code patches instead to
inform test selection.
Through extensive experiments, this paper demonstrates that \oursystem
outperforms prior work on this problem, and that these improvements
can be attributed to the novel features of \oursystem.
The experiments also demonstrate that the tests generated by
\oursystem are effective at boosting precision of SWE agents that
generate patches from issues.
Overall, we hope that by automatically generating reproduction tests,
we can take the tedium out of test-driven development~(TDD), so more
developers and software projects can reap the benefits of TDD.
All our generated tests and logs are available in our repository~\cite{eotter_artifact}.